\documentclass[journal]{IEEEtran}

\usepackage{amssymb, amsthm, amsmath, graphicx, cite}
\usepackage[caption=false,font=footnotesize]{subfig}
\usepackage[ruled]{algorithm2e}

\newtheorem{thm}{Theorem}
\newtheorem{cor}{Corollary}
\newtheorem{lem}{Lemma}
\newtheorem{prop}{Proposition}
\newtheorem{defi}{Definition}

\newcommand{\cent}{0}
\newcommand{\FA}{{\sf FA}}
\newcommand{\MD}{{\sf MD}}
\newcommand{\wh}[1]{\widehat{#1}}

\DeclareMathOperator*{\argmin}{\arg\min}

\title{Decision Making in Star Networks \\ with Incorrect Beliefs}
\author{Daewon Seo, Ravi Kiran Raman, and Lav R.~Varshney
	\thanks{ This work was supported in part by the National Science Foundation under grant CCF-1717530. This paper was presented in part at the 2020 IEEE International Symposium on Information Theory \cite{SeoRV2020}.}
	\thanks{ D.~Seo is with the Department of Information and Communication Engineering, DGIST, Daegu 42988, South Korea (e-mail: dwseo@dgist.ac.kr). R.~K.~Raman is with Analog Devices Inc., Boston, MA 02111 USA (e-mail: ravi.raman@analog.com). L.~R.~Varshney is with the Coordinated Science Laboratory and the Department of Electrical and Computer Engineering, University of Illinois at Urbana-Champaign, Urbana, IL 61801 USA (e-mail: varshney@illinois.edu). }
}

\begin{document}

\maketitle

\begin{abstract}
Consider a Bayesian binary decision-making problem in star networks, where local agents make selfish decisions independently, and a fusion agent makes a final decision based on aggregated decisions and its own private signal. In particular, we assume all agents have private beliefs for the true prior probability, based on which they perform Bayesian decision making. We focus on the Bayes risk of the fusion agent and counterintuitively find that incorrect beliefs could achieve a smaller risk than that when agents know the true prior. It is of independent interest for sociotechnical system design that the optimal beliefs of local agents resemble human probability reweighting models from cumulative prospect theory.

We also consider asymptotic characterization of the optimal beliefs and fusion agent's risk in the number of local agents. We find that the optimal risk of the fusion agent converges to zero exponentially fast as the number of local agents grows. Furthermore, having an identical constant belief is asymptotically optimal in the sense of the risk exponent. For additive Gaussian noise, the optimal belief turns out to be a simple function of only error costs and the risk exponent can be explicitly characterized.
\end{abstract}


\section{Introduction} \label{sec:intro}
Consider a committee where members convey their findings to a chairperson that makes a final decision. Each member has a private assessment of the issue under discussion and presents a finding to the chairperson. The chairperson makes a decision based on his/her own assessment and the findings presented by non-chair members. In such a scenario, it is common to assume members behave selfishly for their own benefit, which is contrary to strategic behavior \cite{Austen-SmithB1996}. Here, selfishness means optimizing one's own performance without consideration of other members, whereas strategic behavior means decision making that optimizes the overall performance of the team/group when considering the strategic responses of other members. Also, each member's selfish assessment is often biased as people may have preconceived beliefs.
 
Another example can be found in labeling data by crowdworkers in crowdsourcing platforms \cite{VempatyVV2014, VempatyVKCV2018}, where several crowdworkers label the same data independently. A collector---possibly another crowdworker---aggregates such labels as well as own observation and makes final labels for the data. Like the previous example, crowdworkers' goal is to maximize their own labeling accuracy so that they maintain a positive reputation in the platform, and the collector's goal is of course to maximize the final accuracy. Here, the crowdworkers and collector may be biased to a certain label when existing labeled data that were used for training are biased.

The setting in these problems arises in numerous technology-mediated social choice systems that must be engineered. Consider recommendation systems in e-commerce based on collaborative filtering. Customers select whether to purchase a product for their own use, but a recommendation algorithm collects those selections and combines them with external information to assess the product's usefulness for future customers \cite{SchaferKR2001}. According to the assessment, the algorithm recommends only useful products to new customers. In addition, customers as well as a recommendation algorithm often have bias in internally representing data (perception) of whether the product will be useful or not. Introducing biased perception is the main distinction of our work from existing literature.

\begin{figure}[t]
	\centering
	\includegraphics[width=2.3in]{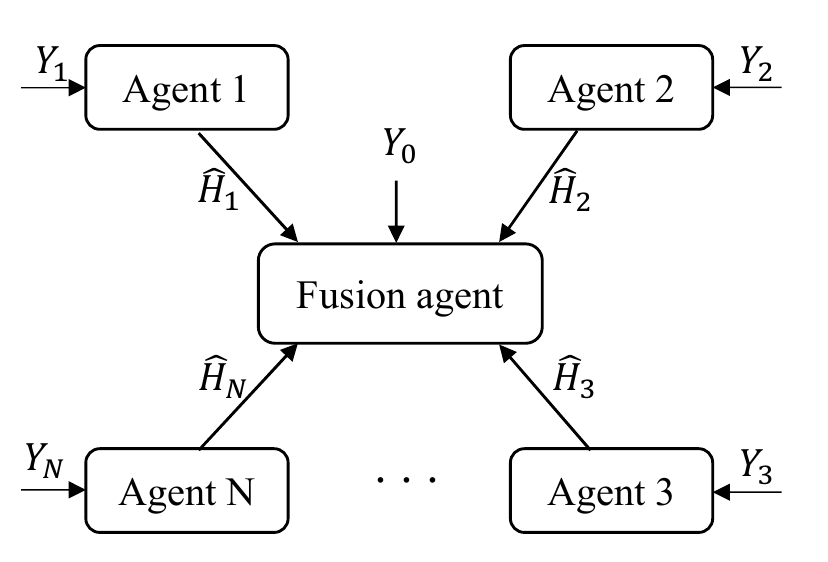}
	\caption{The star network model.}
	\label{fig:star_model}
\end{figure}
Our problem model is depicted in Fig.~\ref{fig:star_model}, but notations will be formalized later. There is a common binary hypothesis posed to a group of $N+1$ agents. There are $N$ local agents that make binary inferences, which we assume selfish, without interacting with others. Such inferences are collected and used to improve the binary choice to be made by the fusion agent. We consider a Bayesian setting, where the hypothesis is randomly drawn according to a prior distribution. However, a novel assumption of our work is that each agent perceives the prior differently from the truth; we say each agent has a \emph{belief} for the true prior. Therefore, each agent makes a Bayes optimal decision according to its own belief instead of the true prior. We suppose the fusion agent makes the final decision in the setup and hence is the most important node. Therefore, our goal is to understand the effect of the biases and the optimal combination of biases that minimizes the Bayes risk of the fusion agent.

The main contribution of this work is twofold. First, we identify that certain combinations of incorrect beliefs of agents outperform the case of all agents making Bayes-optimal decisions based on the true prior, as in Thm.~\ref{thm:derivative_zero}. As finding the globally optimal beliefs is challenging, we suggest a person-by-person optimization (PBPO) algorithm instead. In the setting of additive standard Gaussian noise, the PBPO solution finds globally optimal beliefs obtained by exhaustive search. Furthermore, the optimal local agent has a distorted belief that overweights small prior probabilities and underweights large prior probabilities, and the optimal fusion agent has a belief that is the opposite of the prior, which we refer to as a \emph{contrarian} belief. Although agents in our model need not be humans, considering the local agents as humans reveals a connection to cumulative prospect theory that the shape of the optimal beliefs minimizing the fusion agent's risk is similar to a human perception curve for probability in cumulative prospect theory. So, one can claim that humans are nearly optimal decision makers for information fusion in star networks if heuristics from cumulative prospect theory is permitted.

Second, we consider many-agent asymptotics and show that having an identical belief, differing from the true prior in general, is asymptotically optimal as the number of agents tends to infinity---in this case, our focus is the risk exponent of the fusion agent, that is, the negative logarithm of the risk, normalized by the number of local agents. It turns out that the optimal belief is a constant regardless of the prior, as Thm.~\ref{thm:asymp_optimal_belief} states. In other words, irrespective of the true prior, a certain constant incorrect belief for each agent in the system, asymptotically in the number of agents, minimizes the overall risk achieved by the fusion agent in the decision-making process. In addition, the optimal risk exponent attained in our setting is the best exponent in the strategic star network \cite{Tsitsiklis1988}---the selfishness and misperception asymptotically incur no additional loss in the Bayes risk of the fusion agent, in the risk exponent sense.

There is a large body of related literature. Such inference problems over networks have been an active topic in engineering and statistics for a last few decades. These are so-called \emph{distributed detection} or \emph{inference} \cite{ViswanathanV1997, Varshney1997}, where an agent aggregates decisions from neighbors and makes a decision. We only provide a non-exhaustive list here: Among numerous network structures, the most relevant work to our setting is a star (also called parallel) network, where it is common to consider the setting with or without a fusion agent that aggregates local decisions and makes a final decision \cite[Chap.~3]{Varshney1997}. To ease analysis, the standard setting often assumes that the observations of local agents are independent conditioned on the hypothesis \cite{TsitsiklisA1985}. Variants of the standard setting include when the link between the local agents and the fusion agent may be rate-limited \cite{BergerZV1996} or imperfect \cite{SaligramaAS2006}. Asymptotic analysis of the optimal decision rule is given in \cite{Tsitsiklis1988} as the number of local agents grows, and in \cite{ChamberlandV2003} as the observations are repeated many times while keeping a network structure fixed. Among other network structures, tandem \cite{HellmanC1970, TangPK1991, TayTW2008} and tree \cite{TangPK1993, TayTW2009} networks are also widely studied. For arbitrary graph networks, it is known that the unknown hypothesis can be identifiable by exchanging a local inference with neighbors and updating it over time \cite{AlanyaliVSA2004, RadT2010, AcemogluDLO2011, JadbabaieMST2012, LalithaJS2018}.

Of interest to sociotechnical system designers \cite{SaadGMP2016, EtesamiSMP2018}, unlike machine agents such as sensor nodes, human behavior has bounded rationality and is affected by individual perceptions of the underlying context. To explain why people make irrational decisions under risk, cumulative prospect theory \cite{TverskyK1992} introduces probability reweighting functions. Among many reweighting functions (e.g., \cite{GonzalezW1999} and references therein), the Prelec reweighting function \cite{Prelec1998} satisfies a majority of axioms of cumulative prospect theory, called \emph{compound invariance}, and explains empirical results well. As we will see, the Prelec reweighting functions are similar to the set of beliefs of individual agents that minimize the collective risk in our model. It brings up an application scenario such as human-AI teams, e.g., an AI recommendation algorithm based on decisions made by humans.

However, our model differs from the existing works in two important ways. First, agents do not have knowledge of the true prior, but hold their individual perceptions of it, which we call their beliefs. Therefore, agents make a biased decision based on their own perception. Second, differently from the cooperative decision making of the prior works, the agents in our model make selfish decisions that minimize their individual perceived risk.

Our finding opens a new avenue for improving decision making in networks using non-idealities. Existing literature mainly focuses on the performance in discrete-observation models. For instance, introducing a few Bayes-irrational revealers acting only on their private observations \cite{PeresRSS2017} and adding noise to decision history \cite{LeSB2017} are known to prevent or reduce incorrect detection of the community, which is called an incorrect information cascade \cite{BikhchandaniHW1992, Banerjee1992}. In addition, it is known that diversifying beliefs via stochastic generation \cite{RosasCG2018b} improves an incorrect information cascade. Our work can be thought of as an investigation into the optimal realization of such stochastic beliefs. Further, both works give the same message---a system knowing the true prior has room for further improvement. Our result demonstrates a new case where non-ideality improves performance in a continuous-observation setting. The finding of this work is also different from our previous research \cite{SeoRRGV2019}, where we studied a tandem of agents that have private beliefs and make a decision sequentially. Focusing on the Bayes risk of the last-acting agent of the tandem, we proved in \cite{SeoRRGV2019} that a certain combination of incorrect beliefs achieves a smaller Bayes risk than that with correct beliefs. While the current work also shows a similar conclusion that incorrect beliefs are beneficial in the newly considered star topology, technical details are largely different. For instance, unlike a tandem network, previous (local) decisions are made in parallel without interaction among agents. Therefore, the fusion agent has to take multiple local decisions into account at the same time, which results in non-monotonic behavior in Fig.~2, in stark contrast to the result of \cite{SeoRRGV2019}.

This work was first presented in part in \cite{SeoRV2020}, where the main focus was on the suboptimality of correct beliefs, and our analytic understanding of the optimal beliefs for the many-agent setting was limited. In this work, we further investigate and analytically derive the optimal belief and risk when the number of agents tends to infinity. Also, Gaussian assumption in noise model is relaxed in this work.

The rest of this paper is organized as follows. Sec.~\ref{sec:model} formulates our setting in a star network and provides the required preliminary. Sec.~\ref{sec:finiteN} investigates the setting with a finite number of agents and presents results on optimal beliefs and human agent approximation by cumulative prospect theory. Sec.~\ref{sec:asymp} studies asymptotic optimality when the number of local agents tends to infinity. Sec.~\ref{sec:discussion} concludes the paper.

\section{Problem Description and Preliminary} \label{sec:model}
\subsection{Model}
Consider a star network, depicted in Fig.~\ref{fig:star_model}, consisting of $N$ local agents and a single fusion agent, denoted as agent $\cent$. The underlying hypothesis, $H \in \{0,1\}$, is a binary random variable that follows the prior probability $\mathbb{P}[H = 0] \triangleq \pi_0$ and $\mathbb{P}[H = 1] = 1-\pi_0 \triangleq \bar{\pi}_0$, which is unknown to the agents. Instead of the unknown $\pi_0$, the $i$-th agent, $i \in \{\cent, 1, \ldots, N\}$, believes every agent perceives the same belief as well. The $i$-th agent observes a continuous private signal $Y_i$ with density $f_{Y|H}(y_i|h)$. For the inference problem to be non-trivial, $f_{Y|H}(\cdot|0) \ne f_{Y|H}(\cdot|1)$ and no realization of $Y$ completely determines $H$. The latter condition is formally equivalent to that $f_{Y|H}(\cdot|0), f_{Y|H}(\cdot|1)$ are absolutely continuous \cite{Bartle1966} with respect to each other. In the sequel, numerical evaluations and discussions are with additive Gaussian noise, that is,
\begin{align*}
	Y_i = H + Z_i, \quad i \in \{\cent, 1, \ldots, N\},
\end{align*}
where $Z_i$ is an independent Gaussian noise with zero mean and variance $\sigma^2$.

The $i$-th agent ($i\ne 0$) establishes a selfish binary decision rule based on its own belief $q_i$ and private signal $Y_i$. Hence $\wh{H}_i, i\ne 0$ is a random variable that depends on $(H, q_i, Y_i)$ as well as error costs that will be mentioned. The fusion agent similarly makes a selfish decision, but it also observes previous decisions made by local agents. Therefore, $\wh{H}_0$ is a random variable that depends on $(H, q_0, \wh{H}_1, \ldots, \wh{H}_N)$ as well as error costs. Since $H$ and $\wh{H}_i$ are both binary, two error events can be defined---false alarm (or type I error, i.e., choosing $\wh{H}_i=1$ when $H=0$), and missed detection (or type II error, i.e., choosing $\wh{H}_i=0$ when $H=1$). We assume correct decisions incur no cost. Costs for false alarm and missed detection are denoted by $c_{\FA}$ and $c_{\MD}$, respectively. In addition, we assume that all agents agree on the costs.

Agents are Bayes-rational under $q_i$ and so make decisions that minimize their \emph{perceived} Bayes risk, 
\begin{align}
	R_{i,[i]} \triangleq c_{\FA}q_i p_{\wh{H}_i|H}(1|0)_{[i]} + c_{\MD} (1-q_i) p_{\wh{H}_i|H}(0|1)_{[i]}, \label{eq:perceived_risk}
\end{align}
where $p_{\wh{H}_i|H}(1|0)_{[i]}, p_{\wh{H}_i|H}(0|1)_{[i]}$ are the false alarm and missed detection probabilities ``seen'' by the $i$-th agent as if $q_i$ is the true prior. As the $i$-th agent believes $q_i$ is the truth, it computes the risk to be minimized based only on $q_i$, not on $(\pi_0, \{q_{j}\}_{j\ne i})$. In other words, $R_{i,[i]}$ is a function of the $i$-th agent's decision rule based on $q_i$. In particular, $R_{\cent, [\cent]}$ is the fusion agent's perceived risk. The fusion agent thinks $q_0$ is the true prior and that the local agents share the same belief. Thus, it interprets local decisions as if they are based on $q_\cent$. Formal calculation of the error probabilities are given in Sec.~\ref{subsec:belief_update}. Risks and error probabilities that do not include a subscript $[i]$ are the \emph{true} quantities computed by an external oracle that has knowledge of $(\pi_0, q_\cent, q_1, \ldots, q_N)$,
\begin{align}
	R_{i} \triangleq c_{\FA}\pi_0 p_{\wh{H}_i|H}(1|0) + c_{\MD} \bar{\pi}_0 p_{\wh{H}_i|H}(0|1). \label{eq:true_risk}
\end{align}

We suppose that the fusion agent is the most important node as its decision is final in the setup. Then, the natural question is under what $(q_0, q_1, \ldots, q_N)$ the true Bayes risk of the fusion agent, $R_0$, is minimized. To simplify analysis, two cases when the number of agents is finite and grows without bound are considered separately.

We do not consider a scenario where the fusion agent infers a belief of a local agent and tweaks $q_\cent$ for a better decision. It is because inferring $q_i$, a continuous value, from a single binary decision $\wh{H}_i$ inevitably incurs errors in inference such that $|\wh{q}_i - q_i| > 0$ in probability, cf.~compressing and decompressing a uniform random variable into $1$ bit \cite{CoverT1991}. For the belief inference to be accurate, the fusion agent must have a number of decisions per agent, which is also often infeasible in particular when the decisions are from anonymous agents.

\subsection{Cumulative Prospect Theory}
Cumulative prospect theory is a widely accepted theory of biased perception by people \cite{TverskyK1992}. According to it, people behave irrationally under risk, e.g., casinos, lotteries, and financial markets, because the human perception of probabilities is distorted. So, cumulative prospect theory considers humans as Bayesian decision makers with prior being reweighted. Here, we review one of the widely used probability reweighting functions in cumulative prospect theory: the Prelec reweighting function. Unlike other empirical reweighting functions \cite{GonzalezW1999}, it is obtained from a set of axioms called \emph{compound invariance}. In addition, it has been found to explain several observed irrational traits of humans and is defined as follows.
\begin{defi}[\!\!\cite{Prelec1998}] \label{def:prelec}
	For $\alpha, \beta > 0$, the Prelec reweighting function $w:[0,1] \mapsto [0,1]$ is 
	\begin{align*}
	w(p;\alpha,\beta) = \exp(-\beta(-\log p)^{\alpha}).
	\end{align*}
\end{defi}
The function satisfies several properties such as: 
\begin{enumerate}
	\item $w(0; \alpha, \beta) = 0$, $w(1; \alpha, \beta) = 1$, and $w(p;\alpha, \beta)$ is strictly increasing in $p$.
	\item When $\alpha < 1$, it spans a class of curves that overweight small probabilities and underweight large probabilities. Similarly, $\alpha > 1$ spans a class of the opposite shape curves, i.e., underweight small probabilities and overweight large ones.
\end{enumerate}
We will approximate optimal beliefs using the Prelec function and discuss inferring from human decisions.

\subsection{Definition and Notation}
To simplify notation, we use $x^N = (x_1, \ldots, x_N)$ to denote a tuple of length $N$, and $x_{-i}^N = (x_1, \ldots, x_{i-1}, x_{i+1}, \ldots, x_N)$ to denote a tuple excluding the $i$-th element. All logarithms are natural logarithms. We use $p, f$ to denote probability mass and density functions, respectively. $Q(x)$ is the complementary cumulative distribution function of the standard Gaussian,
\begin{align*}
	Q(x) = \int_x^{\infty} \phi(t;0,1) dt,
\end{align*}
where $\phi(\cdot; \mu, \sigma^2)$ is the probability density function of the Gaussian random variable with mean $\mu$ and variance $\sigma^2$. We say the private observation is sub-Gaussian if $f_{Y|H}(\cdot|h)$ are sub-Gaussian \cite{Vershynin2018}, i.e.,  for all $s \in \mathbb{R}, h \in \{0,1\}$, there is a variance proxy $\sigma^2$ such that
\begin{align*}
	\mathbb{E}_{Y|H=h}[\exp(s(Y-\mathbb{E}_{Y|H=h}[Y])] \le \exp\left( \frac{\sigma^2 s^2}{2} \right).
\end{align*}
Also, we use $C(\cdot, \cdot)$ to denote the Chernoff information between two probability distributions over a discrete space $\mathcal{X}$ \cite{Chernoff1952},
\begin{align*}
	C(p_1, p_2) = - \min_{s \in [0,1]} \log \left( \sum_{x \in \mathcal{X}} p_1^s(x) p_2^{1-s}(x) \right).
\end{align*}

\section{Optimal Beliefs for Finite $N$} \label{sec:finiteN}
\subsection{Belief Update} \label{subsec:belief_update}
Before discussing the optimal decision rules of agents, let us consider the standard case when agents know the true prior. Since agents know $\pi_0$, the $i$-th agent can compute the true risk and wishes to minimize the risk \eqref{eq:true_risk} by optimizing the decision rule. Any decision rule, in general, is admissible, but the likelihood ratio test (LRT) minimizes the risk $R_i$ \cite{VanTrees1968}. That is, if agent $i \in \{1,\ldots, N\}$ knows the true prior, the optimal decision rule is as follows.
\begin{align}
	\frac{f_{Y|H}(y_i|1)}{f_{Y|H}(y_i|0)} \underset{\wh{H}_i = 0}{\overset{\wh{H}_i = 1}{\gtrless}} \frac{c_{\FA} \pi_0}{c_{\MD}(1-\pi_0)},
\end{align}
where $\underset{\wh{H}_i = 0}{\overset{\wh{H}_i = 1}{\gtrless}}$ indicates that $\wh{H}_i=1$ if the likelihood ratio (the left side) is greater than the decision threshold (the right side), and declares $\wh{H}_i=0$ if the likelihood ratio is smaller than the threshold. As the distribution of $y$ is continuous, ties occur with measure zero and can be resolved arbitrarily without affecting the risk.

Returning to our case when agent $i$ believes that $q_i$ is the true prior, the LRT decision rule that minimizes the perceived risk $R_{i, [i]}$ in \eqref{eq:perceived_risk} reduces to
\begin{align}
\frac{f_{Y|H}(y_i|1)}{f_{Y|H}(y_i|0)} \underset{\wh{H}_i = 0}{\overset{\wh{H}_i = 1}{\gtrless}} \frac{c_{\FA}q_i}{c_{\MD}(1-q_i)} \triangleq \lambda_i = \lambda(q_i). \label{eq:likelihood_test_distributed}
\end{align}

Noting that when the noise is Gaussian, $f_{Y|H}(y_i|h)$ is Gaussian with mean $h$ and variance $\sigma^2$, \eqref{eq:likelihood_test_distributed} can be simplified to
\begin{align}
y_i \underset{\wh{H}_i = 0}{\overset{\wh{H}_i = 1}{\gtrless}} \lambda(q_i) = \frac{1}{2} + \sigma^2 \log\left( \frac{c_{\FA}q_i}{c_{\MD}(1-q_i)} \right). \label{eq:decision_threshold}
\end{align}
In this case, the conditional error probabilities for local agents have a closed-form expression in terms of the $Q$-function,
\begin{subequations} \label{eq:q_function}
	\begin{align}
		p_{\wh{H}_i|H}(1|0) &= \int_{\lambda_i}^{\infty} \phi(t;0, \sigma^2) dt = Q \left( \frac{\lambda_i}{\sigma} \right), \\
		p_{\wh{H}_i|H}(0|1) &= \int_{-\infty}^{\lambda_i} \phi(t;1, \sigma^2) dt = 1-Q\left( \frac{\lambda_i-1}{\sigma} \right).
	\end{align}
\end{subequations}

The fusion agent with belief $q_\cent$ has access to all decisions made by local agents, so its LRT, given $(y_\cent, \wh{h}_1, \ldots, \wh{h}_N)$ is
\begin{align*}
	\frac{f_{Y, \wh{H}^N|H}(y_\cent, \wh{h}^N|1)}{f_{Y, \wh{H}^N |H}(y_\cent, \wh{h}^N|0)} \underset{\wh{H}_\cent = 0}{\overset{\wh{H}_\cent = 1}{\gtrless}} \frac{c_{\FA} q_\cent}{c_{\MD}(1-q_\cent)}.
\end{align*}
Since $Y_\cent, \wh{H}_1, \ldots, \wh{H}_N$ are independent conditioned on $H$,
\begin{align*}
	f_{Y, \wh{H}^N |H}(y_\cent, \wh{h}^N|h) = f_{Y|H}(y_\cent|h)\prod_{i=1}^N  p_{\wh{H}_i|H}(\wh{h}_i|h).
\end{align*}
Here $p_{\wh{H}_i|H}$ is a function of $q_i$ only. However, the fusion agent believes $q_\cent$ is the true prior that other agents use too. So, the fusion agent computes $p_{\wh{H}_i|H}$ as if all local agents performed hypothesis testing \eqref{eq:likelihood_test_distributed} with $q_\cent$. It leads to the following LRT that the fusion agent actually performs\footnote{Again, the subscript $[\cent]$ denotes quantities that the fusion agent ``thinks'', i.e., computation is based on $q_\cent$.}
\begin{align}
\frac{f_{Y|H}(y_\cent|1)}{f_{Y|H}(y_\cent|0)} \underset{\wh{H}_\cent = 0}{\overset{\wh{H}_\cent = 1}{\gtrless}} \frac{c_{\FA} q_\cent}{c_{\MD}(1-q_\cent)} \prod_{i=1}^{N} \frac{p_{\wh{H}_i|H}(\wh{h}_i|0)_{[\cent]}}{p_{\wh{H}_i|H}(\wh{h}_i|1)_{[\cent]}}. \label{eq:LLR_cent_rearrange}
\end{align}
Since $x/(1-x)$ is monotonically increasing in $x \in (0,1)$, we can translate \eqref{eq:LLR_cent_rearrange} into a new LRT with updated belief $q_\cent'$,
\begin{align}
\frac{f_{Y|H}(y_\cent|1)}{f_{Y|H}(y_\cent|0)} \underset{\wh{H}_\cent = 0}{\overset{\wh{H}_\cent = 1}{\gtrless}} \frac{c_{\FA} q_\cent'}{c_{\MD}(1-q_\cent')}, \label{eq:LLR_cent_new_belief}
\end{align}
where $q_\cent'$ satisfies
\begin{align}
\frac{q_\cent'}{1-q_\cent'} = \frac{q_\cent}{1-q_\cent} \prod_{i=1}^{N} \frac{p_{\wh{H}_i|H}(\wh{h}_i|0)_{[\cent]}}{p_{\wh{H}_i|H}(\wh{h}_i|1)_{[\cent]}}. \label{eq:belief_update}
\end{align}

Finally, the true Bayes risk of the fusion agent is
\begin{align}
R_\cent = c_{\FA} \pi_\cent p_{\wh{H}_\cent|H}(1|0) + c_{\MD}\bar{\pi}_0 p_{\wh{H}_\cent|H}(0|1), \label{eq:R_N}
\end{align}
with
\begin{align*}
	p_{\wh{H}_\cent|H}(\wh{h}_\cent|h) = \sum_{\wh{h}^N} p_{\wh{H}^N, \wh{H}_\cent|H}(\wh{h}^N, \wh{h}_\cent|h).
\end{align*}

\begin{figure}[t]
	\centering
	\subfloat[$N=2$]{\includegraphics[width=1.82in]{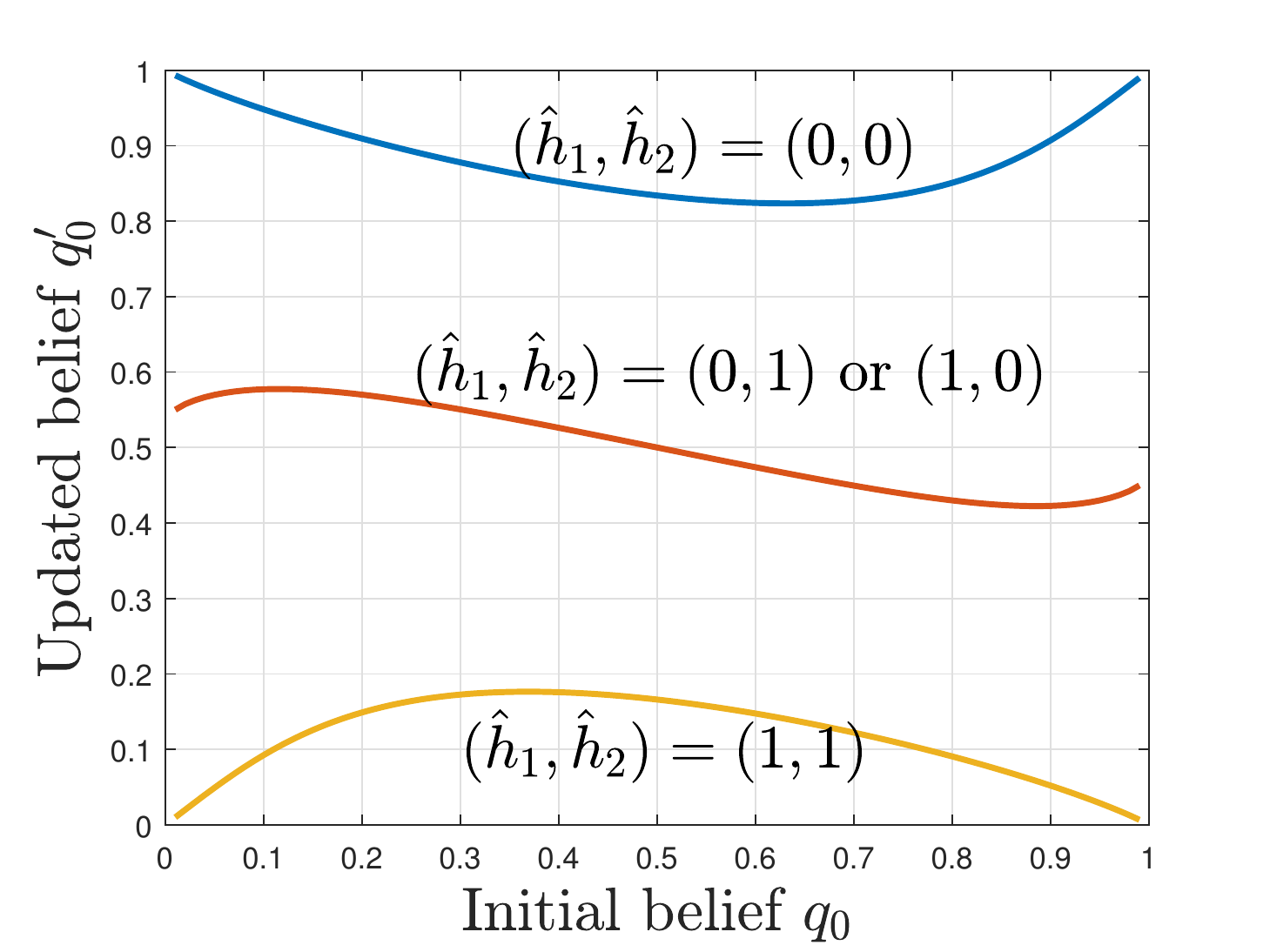}}
	\subfloat[$N=3$]{\includegraphics[width=1.82in]{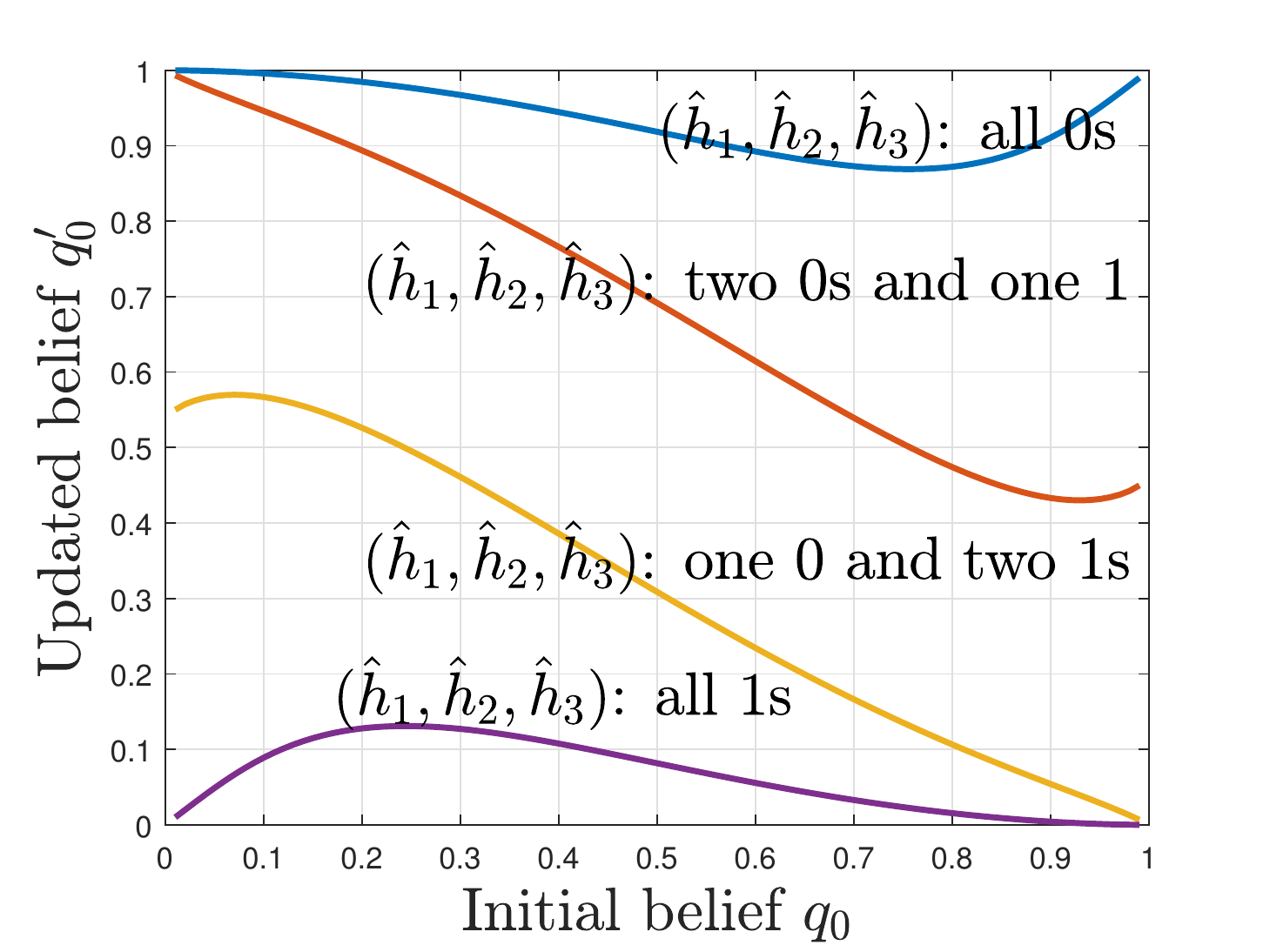}}
	\caption{Updated belief \eqref{eq:belief_update} for all possible local decisions. Additive standard Gaussian noise is assumed.}
	\label{fig:belief_update_N=2}
\end{figure}
Let us consider numerical evaluations with $N=2,3$, assuming additive standard Gaussian noise. As stated, the fusion agent adopts the new LRT based on the updated belief $q_\cent'$ as in \eqref{eq:LLR_cent_new_belief}. Fig.~\ref{fig:belief_update_N=2} depicts the updated belief $q_\cent'$ in \eqref{eq:belief_update} for possible local decisions. The curves indicate how observing local decisions changes the fusion agent's initial belief. We can see that $q_\cent$ significantly changes when observed decisions significantly differ from what the fusion agent expects. For example in Fig.~\ref{fig:belief_update_N=2}(b), when $q_\cent$ is small, the fusion agent believes $H$ is highly likely to be $1$. However, observing $(\wh{h}_1, \wh{h}_2, \wh{h}_3)=(0,0,0)$, its updated belief becomes close to $1$ implying that it now believes $H$ is highly likely to be $0$. On the other hand, observing $(\wh{h}_1, \wh{h}_2, \wh{h}_3)=(1,1,1)$ it confirms and further enhances the small $q_\cent$, so $q_\cent' < q_\cent$ after observing $(\wh{h}_1, \wh{h}_2, \wh{h}_3)=(1,1,1)$.

Note that the updated belief curves for each set of prior decisions are not monotonic in $q_\cent$, which is different from what is observed for a tandem network \cite{SeoRRGV2019}. The belief update in a tandem network is indeed the same as \eqref{eq:belief_update} with $N=1$ at each update step, where it was proven that the updated belief is always monotonically increasing in $q_\cent$ \cite[Fig.~2 and Thm.~3]{SeoRRGV2019}. However, it is no longer true in the star network where multiple local decisions ($N \ge 2$) should be considered simultaneously, as illustrated in Fig.~\ref{fig:belief_update_N=2}.

By further looking into \eqref{eq:belief_update}, we can see why such non-monotonicity occurs. Let us define $\Delta(q_\cent) \triangleq q_\cent'(q_\cent) - q_\cent$ as the amount of ``surprise'' in the local decisions, clearly a function of $q_\cent$. Using $\Delta(q_\cent)$, we can separately characterize the effect of initial belief and observed decisions in the updated belief. Note that the updated belief $q_\cent'(q_\cent)$ is monotonically increasing if and only if $\tfrac{\partial q_\cent'}{\partial q_\cent} = 1 + \tfrac{\partial \Delta}{\partial q_\cent} > 0$. Since $q_0'$ is non-monotonic, we can conclude that the surprise $\Delta(q_\cent)$ could change more rapidly below $-1$ as $q_\cent$ varies only when $N \ge 2$. This can be directly observed from \eqref{eq:belief_update}: Note that $q_\cent/(1-q_\cent)$ is increasing in $q_\cent$, whereas $p_{\wh{H}_i|H}(\wh{h}_i|0)_{[\cent]} / p_{\wh{H}_i|H}(\wh{h}_i|1)_{[\cent]}$ is decreasing in $q_\cent$ for both $\wh{h}_i=0,1$. So, the non-monotonicity occurs when the last multiplicative factors in the right side of \eqref{eq:belief_update} are large enough to counter the increment of $q_\cent/(1-q_\cent)$ term. This is the case for Gaussian noise. In such cases, the change in the surprise (i.e., the surprise from the local decisions) outweighs the perceived belief change if $N \ge 2$, whereas the change in initial perception (i.e., $q_\cent$) always dominates in tandem networks \cite{SeoRRGV2019}.

\subsection{Optimal Beliefs}
Following the LRTs \eqref{eq:likelihood_test_distributed} and \eqref{eq:LLR_cent_rearrange}, the local agents decide, and the fusion agent's risk $R_\cent$ can be computed by \eqref{eq:R_N}. Note that $R_\cent$ is a function of $(q_\cent, q_1, \ldots, q_N)$ for given $\pi_0$ and costs. One might think that $R_\cent$ achieves its minimum when each agent knows the true prior, i.e., when $q_\cent = q_1=\cdots=q_N = \pi_0$, since the local agents make the best local decisions and the fusion agent does not misunderstand the context the local decisions were made. However, we will see that this is false.

Recall local decisions are independent conditioned on $H$, which implies $P_{\wh{H}_\cent|H}(\wh{h}_\cent|h)$ in \eqref{eq:R_N} can be rewritten as
\begin{align*}
	p_{\wh{H}_\cent|H}(\wh{h}_\cent|h) &= \sum_{\wh{h}^N} \left( \prod_{i=1}^N p_{\wh{H}_i|H}(\wh{h}_i|h) \right) p_{\wh{H}_\cent|H, \wh{H}^N}(\wh{h}_\cent|h, \wh{h}^N).
\end{align*}
Then, \eqref{eq:R_N} can be expressed as
\begin{align}
	&R_\cent = c_{\FA}\pi_0 \sum_{\wh{h}^N} \left( \prod_{i=1}^N p_{\wh{H}_i|H}(\wh{h}_i|0) \right) p_{\wh{H}_\cent|H, \wh{H}^N}(1|0, \wh{h}^N) \nonumber \\
	&~~ + c_{\MD}\bar{\pi}_0 \sum_{\wh{h}^N} \left( \prod_{i=1}^N p_{\wh{H}_i|H}(\wh{h}_i|1) \right) p_{\wh{H}_\cent|H, \wh{H}^N}(0|1, \wh{h}^N). \label{eq:risk_expression}
\end{align}

The next theorem gives a condition that the optimal beliefs achieving the smallest risk at the fusion agent must satisfy. It shows that making decisions based on the true prior is, in general, suboptimal, i.e., the optimal belief $q_i^* \ne \pi_0$ in general. Before proceeding, let us define conditional error probabilities for notational brevity. For $j \in \{1, \ldots, N\}$,
\begin{subequations} \label{eq:def_p_FA_p_MD}
\begin{align}
	&p_\FA(\wh{h}_j=h) \nonumber \\
	&= \sum_{\wh{h}_{-j}^N} \left( \prod_{i \ne j} p_{\wh{H}_i|H}(\wh{h}_i|0) \right) p_{\wh{H}_\cent|H,\wh{H}_{-j}^N,\wh{H}_j}(1|0, \wh{h}_{-j}^N, h), \\
	&p_\MD(\wh{h}_j=h) \nonumber \\
	&= \sum_{\wh{h}_{-j}^N} \left( \prod_{i \ne j} p_{\wh{H}_i|H}(\wh{h}_i|1) \right) p_{\wh{H}_\cent|H,\wh{H}_{-j}^N,\wh{H}_j}(0|1, \wh{h}_{-j}^N, h).
\end{align}
\end{subequations}
By the sum over all decisions except for the $j$-th agent's decision, $p_\FA(\wh{h}_j=h)$ (or $p_\MD(\wh{h}_j=h)$) is the false alarm (or missed detection) probability of the fusion agent when the $j$-th agent declares $\wh{h}_j=h$.

\begin{thm} \label{thm:derivative_zero}
Let $(q_\cent^*, q_1^*, \ldots, q_N^*)$ be the optimal belief tuple that minimizes $R_\cent$. Then, it must satisfy
\begin{align}
\frac{q_j}{1-q_j} = \frac{\pi_0}{1-\pi_0} \frac{p_\FA(\wh{h}_j=1) - p_\FA(\wh{h}_j=0)}{p_\MD(\wh{h}_j=0) - p_\MD(\wh{h}_j=1)} \label{eq:diff_zero}
\end{align}
for all $j \in \{1, \ldots, N\}$.
\end{thm}
\begin{IEEEproof}
Differentiating \eqref{eq:risk_expression} with respect to decision threshold $\lambda_j$ and rearranging terms prove the claim. Details are in App.~\ref{app:derivative_zero}.
\end{IEEEproof}

Note that the fusion agent is more likely to declare $\wh{h}_\cent=1$ upon observing $\wh{h}_j=1$ than before observing it. So, the last factor in \eqref{eq:diff_zero} is the ratio of the change in false alarm to the change in missed detection between $\wh{h}_j=0, 1$. Also the right side of \eqref{eq:diff_zero} is independent of $q_j$ since $p_\FA(\wh{h}_j=h), p_\MD(\wh{h}_j=h)$ are the false alarm and missed detection probabilities of the fusion agent provided that $\wh{h}_j=h$. Hence, \eqref{eq:diff_zero} can be thought of as a balance condition that the optimal initial beliefs must satisfy for error probability changes. As the value  $\frac{p_\FA(\wh{h}_j=1) - p_\FA(\wh{h}_j=0)}{p_\MD(\wh{h}_j=0) - p_\MD(\wh{h}_j=1)}$ is not $1$ in general, $q_i^* \neq \pi_0$ in general.

\begin{figure}[t]
	\centering
	\includegraphics[width=3.0in]{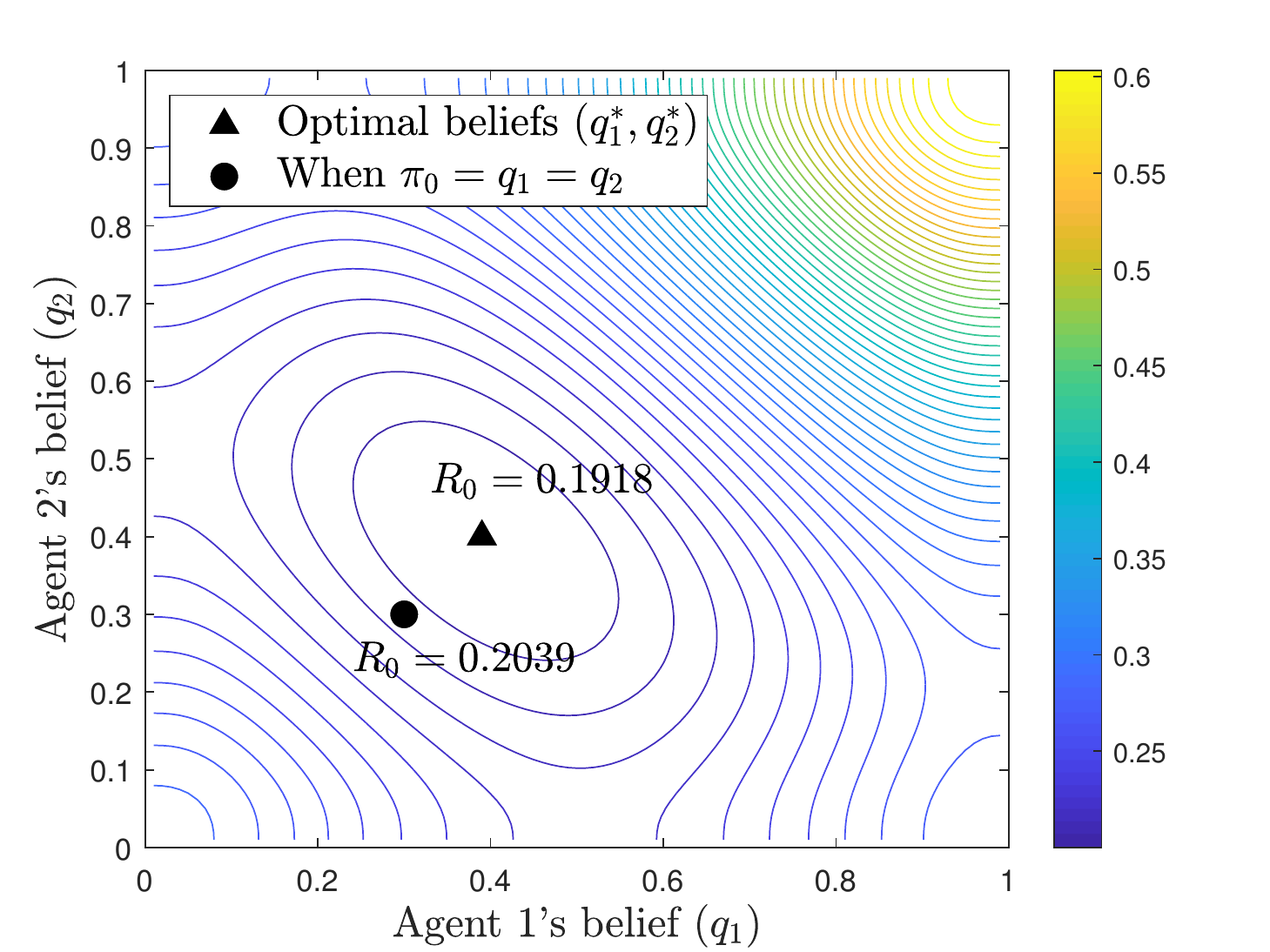}
	\caption{Risk contour for $N=2$ at $\pi_0 = 0.3$, $q_\cent = 0.7372$, $c_{\FA} = c_{\MD}=1$. Additive standard Gaussian noise is assumed. The smallest $R_0=0.1918$ is attained at $q_0=0.7372, q_1=q_2=0.3960$.}
	\label{fig:risk_contour}
\end{figure}
\begin{figure}[t]
	\centering
	\subfloat[$c_{\FA} = c_{\MD}=1$]{\includegraphics[width=2.4in]{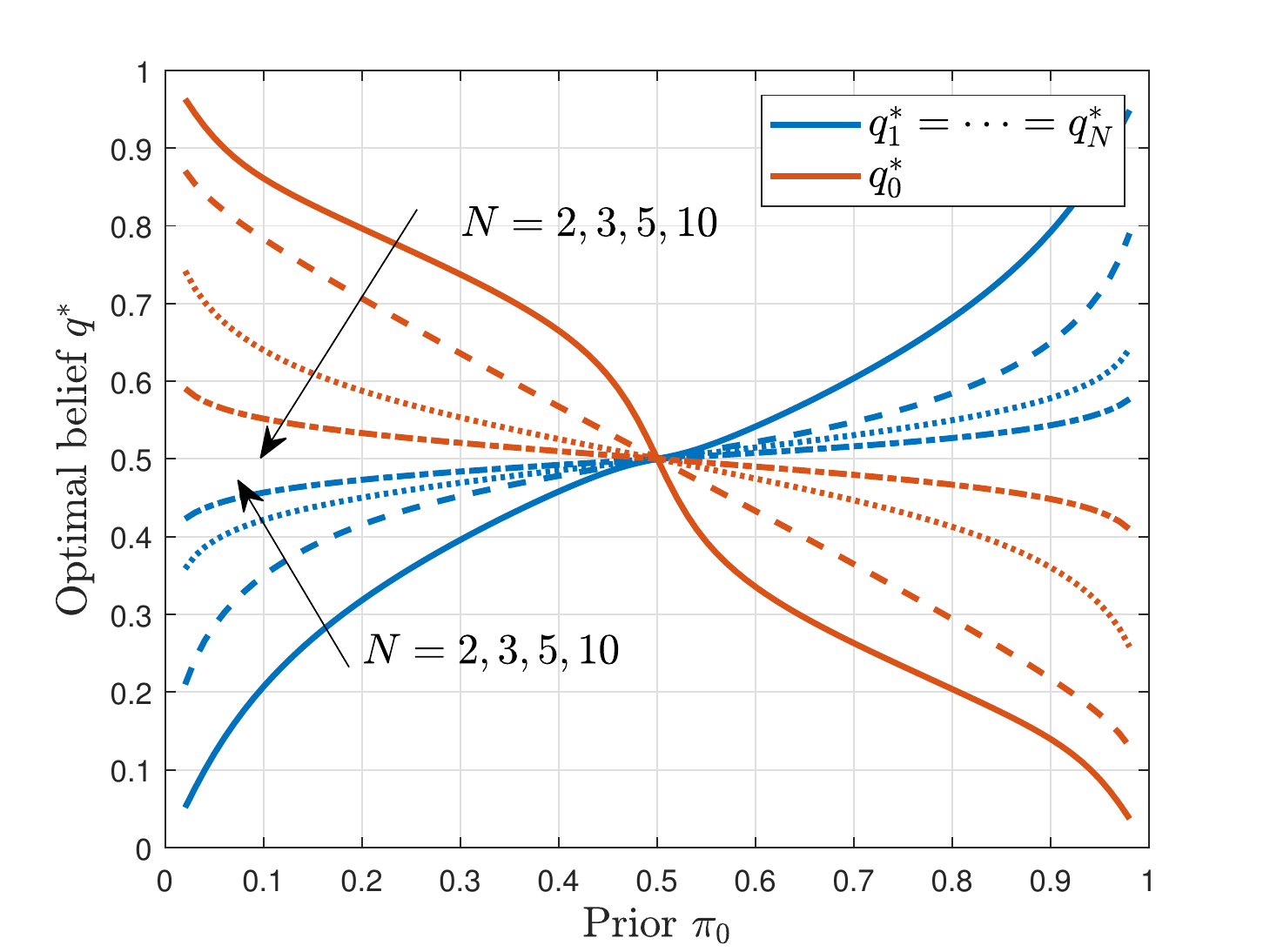}} \\
	\subfloat[$c_{\FA} = 1$, $c_{\MD}=2$]{\includegraphics[width=2.4in]{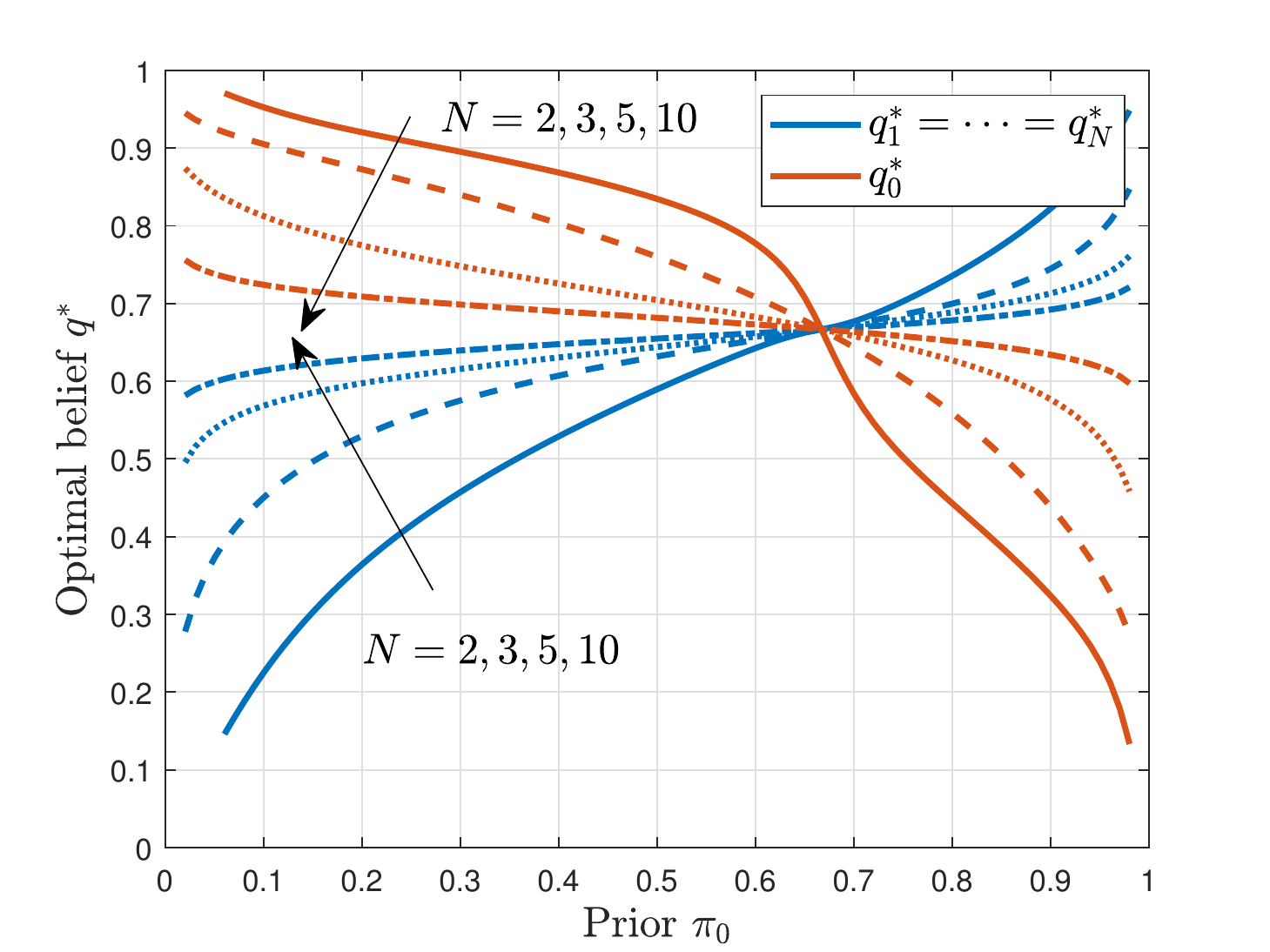}}
	\caption{Optimal beliefs that minimize $R_\cent$ for several numbers of local agents. Additive standard Gaussian noise is assumed. The curves for $N=2,3$ (solid and dashed, respectively) are found by exhaustive search to obtain the global minimum, whereas curves for $N=5,10$ (dotted and dash-dot, respectively) are drawn by assuming $q_1 = q_2 = \cdots = q_N$, i.e., possibly a local minimum. Arrows indicate ``as $N$ grows''.}
	\label{fig:opt_belief_trend}
\end{figure}
Fig.~\ref{fig:risk_contour} depicts the optimal beliefs for additive standard Gaussian noise with $N=2$, $c_{\FA} = c_{\MD} = 1$, and $\pi_0 = 0.3$. In this case, the optimal beliefs are $q_\cent = 0.7372$, $q_1 = q_2 = 0.3960$, as shown in Fig.~\ref{fig:opt_belief_trend}(a), and resulting minimal Bayes risk (triangle) $R_\cent = 0.1918$. Clearly, prior-aware local agents $\pi_0 = q_1 = q_2$ underperform with $R_\cent = 0.2039$ (circle). Also remark that when $\pi_0 = q_1 = q_2 = q_\cent$, not shown in Fig.~\ref{fig:risk_contour}, $R_\cent = 0.1976$ is attained, which is strictly worse than the optimal risk.

This result can be intuitively understood as distortion caused by decision making. As the decision-making process can be thought of as quantization, what the fusion agent aggregates are quantized versions of continuous signals $Y_i$. Noting quantization inevitably incurs distortion, the fusion agent attains the best performance when local agents' distortion and own distorted understanding, caused by the local agents' bias and own bias respectively, fit well.

Fig.~\ref{fig:opt_belief_trend} depicts the optimal beliefs\footnote{However, curves for $N=5,10$ are drawn with the assumption of identical local beliefs, i.e., possibly a local minimum.} for $N=2,3,5,10$ as $\pi_0$ changes. Two observations can be made: First, the optimal beliefs come closer to $\frac{c_{\MD}}{ c_{\FA}+c_{\MD} }$ as $N$ grows for the entire range of $\pi_0$. It suggests that setting $q_i = \frac{c_{\MD}}{ c_{\FA}+c_{\MD} }$ for all $i \in\{\cent, 1,\ldots, N\}$ would be asymptotically optimal as $N$ grows. If this is true, the optimal beliefs are universally optimal, i.e., do not depend on $\pi_0$. We rigorously revisit this in Thm.~\ref{thm:asymp_optimal_belief} in the risk exponent sense. Second, the optimal local belief overweights small probabilities and underweights large probabilities so it tends to be more neutral than the true prior. In other words, to achieve the least Bayes risk of the fusion agent, the local agents have to think the hypothesis is fairer than the truth. However, the optimal local belief still overweights or underweights the horizontal line $\frac{c_{\MD}}{ c_{\FA}+c_{\MD} }$. To compensate for this bias, the fusion agent has to hold a contrarian belief that is contrary to the true prior, i.e., believes $H=0$ more likely when $H=1$ is, in fact, more likely, and vice versa.

Since a local decision is Bayes optimal only when the decision is with the accurate knowledge of the prior, Thm.~\ref{thm:derivative_zero} indicates that the risk of the fusion agent is minimized only at the expense of risks of the local agents. However, because each agent minimized \emph{perceived} individual Bayes risk, the local agents do not realize that they sacrifice their risk for the fusion agent. Thus, if an organizer of the decision-making system intentionally provides the agents with appropriately incorrect knowledge about the prior and makes the agents believe it, local agents maximize the system performance without being aware of their loss in risk. Recalling the labeling example with crowdworkers, we can see that misinforming the crowdworkers and collector of incorrect prior, for instance by providing biased training data, is necessary for the smallest final risk. However, it is notable that a data provider must know the true prior as the optimal biased perception depends on the true prior.

The global optimization problem for $R_\cent$ over $(q_\cent, q_1, \ldots, q_N)$ belongs to neither a convex class nor any analytically solvable classes, as far as we know. A popular numerical approach for this is the person-by-person optimization (PBPO) that optimizes only one variable at a time with other variables being fixed, e.g., \cite{HoballahV1989, TangPK1991b}. The next lemma proves that $R_\cent$ is coordinate-wise convex, which enables us to apply the PBPO to our setting.
\begin{lem} \label{lem:convexity}
	$R_\cent$ is strictly convex in $p_{\wh{H}_j|H}(1|0), j \in \{\cent, 1,\ldots, N\}$ when other quantities are fixed.
\end{lem}
\begin{IEEEproof}
Provided in App.~\ref{app:convexity}.
\end{IEEEproof}

\begin{algorithm}
	\caption{PBPO Algorithm} \label{alg:PBPO}
	\textbf{Hyperparameter:} Step size $\Delta$, stopping criterion $\epsilon > 0$ \\
	\textbf{Output:} PBPO optimal belief $(q_\cent, \ldots, q_N)$ \\
	
	\vspace{0.1in}
	
	Initialize $q_i$, $i= \cent, 1,\ldots, N$ arbitrarily\;
	
	\While{True}{		
		\texttt{*Update $q_i$*} \\
		\For{$i=0, \ldots, N$}{
			$\text{Risk}_+ \gets R_0(q_0', q_1', \ldots, q_{i-1}', q_i+\Delta, q_{i+1}, \ldots)$\;
			$\text{Risk}_- \gets R_0(q_0', q_1', \ldots, q_{i-1}', q_i-\Delta, q_{i+1}, \ldots)$\;
			\eIf{$\text{Risk}_+ < \text{Risk}_-$}{
				$q_i' \gets q_i + \Delta$
			}{
				$q_i' \gets q_i - \Delta$
			}
		}
		\vspace{0.1in}
		\texttt{*Stopping criterion*} \\
		\If{$||(q_\cent', \ldots, q_N') - (q_\cent, \ldots, q_N)||_2 \le \epsilon$}{
			Algorithm stops\;
		}
	}
\end{algorithm}
\begin{figure}[t]
	\centering
	\subfloat[Belief]{\includegraphics[width=1.82in]{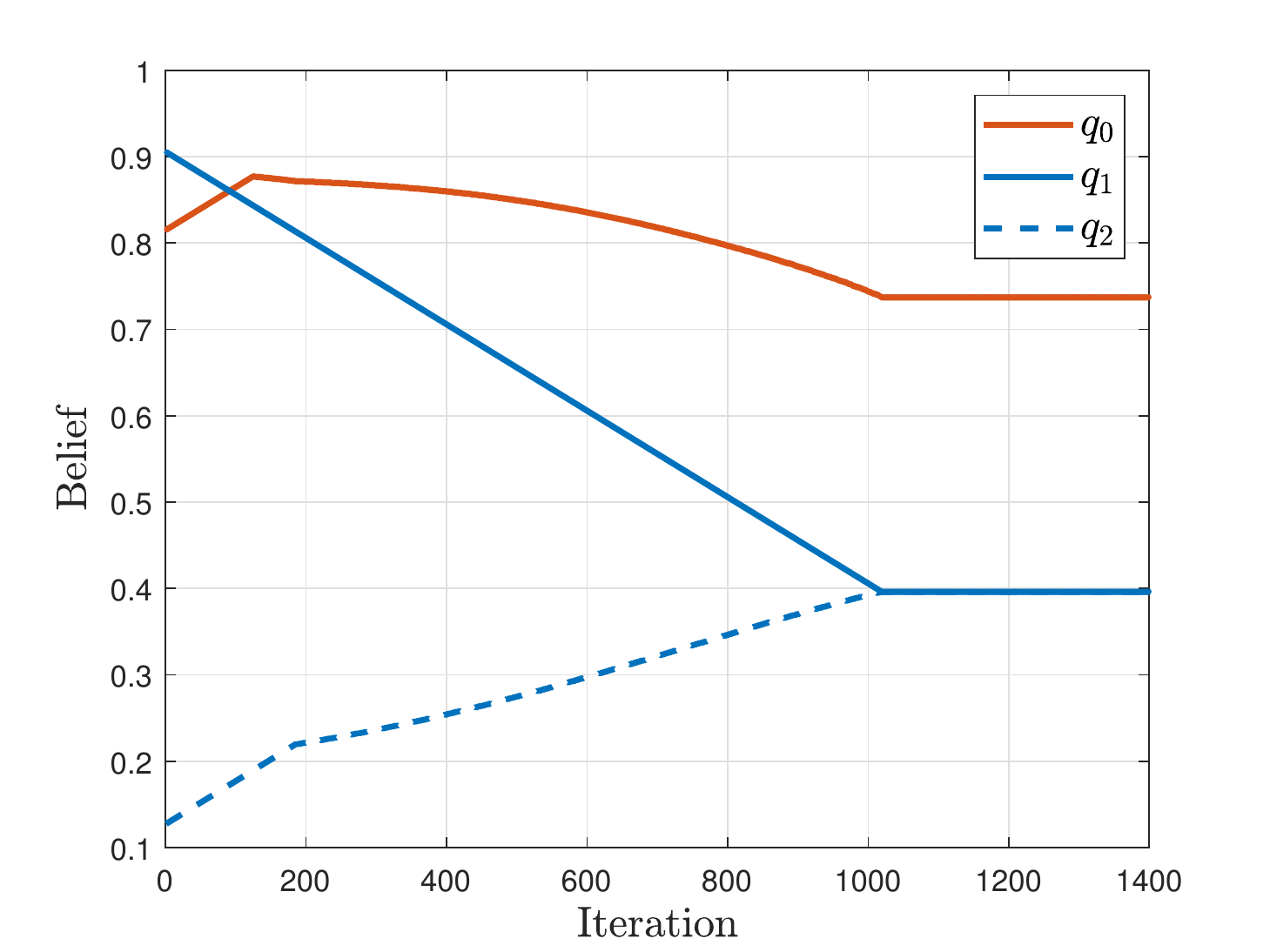}}
	\subfloat[Risk]{\includegraphics[width=1.82in]{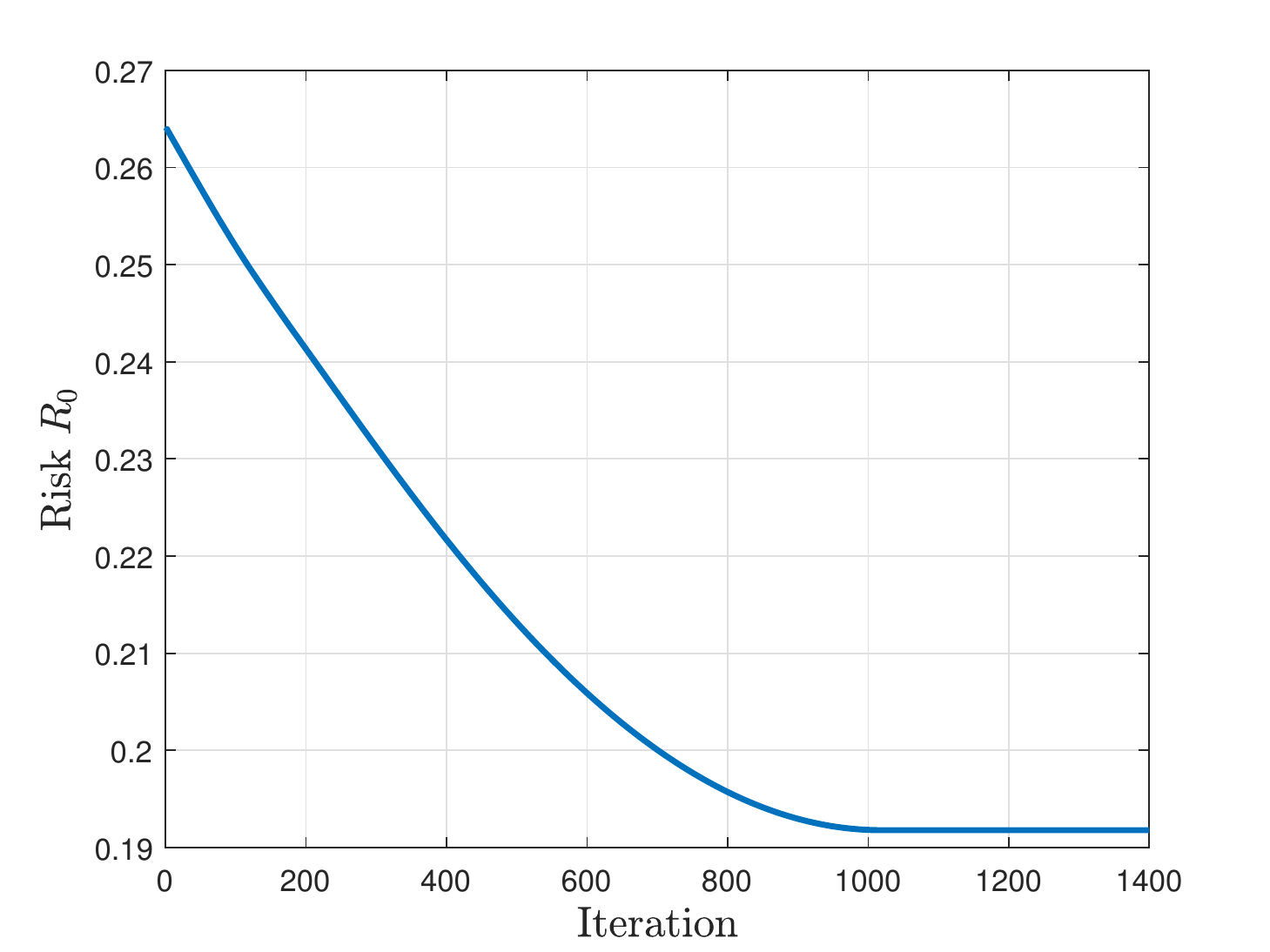}}
	\caption{A sample trajectory of $q_0, q_1, q_2, R_0$ via Algorithm \ref{alg:PBPO}. The setting is the same as Fig.~\ref{fig:risk_contour}: $N=2, \pi_0=0.3, c_\FA=c_\MD=1$, additive standard Gaussian noise. The algorithm with $\Delta=0.0005$ converged to $q_0=0.7372$, $q_1=q_2=0.3960$, $R_0=0.1918$: The same result as exhaustive search was obtained.}
	\label{fig:PBPO}
\end{figure}

Therefore, any convex optimization algorithm with respect to $\{p_{\wh{H}_j|H}(1|0)\}_j$ numerically finds the PBPO solution $\{p_{\wh{H}_j|H}(1|0)\}_j$, which in turn results in the PBPO solution $(q_\cent, q_1, \ldots, q_N)$ since they are in continuous bijection. Alg.~\ref{alg:PBPO} shows a na\"{i}ve PBPO algorithm with Gauss-Seidel update. Throughout the algorithm, the step size is constant.

The algorithm exhibits monotonically decreasing $R_\cent$ over each iteration and $R_\cent$ is bounded below, hence, converges. Once the convergence occurs, there is no decrease in $R_\cent$ along any $i$-th direction; therefore, attains either a local minimum or a saddle point \cite{Bertsekas1982}. Repeating the entire algorithm a number of times with different initializations and selecting the solution that yields the least risk gives the global solution numerically.

A sample trajectory of beliefs and risk during iterations of the algorithm is shown in Fig.~\ref{fig:PBPO}, assuming $\pi_0 = 0.3$ and standard Gaussian noise. With $\Delta=0.0005$ being fixed throughout the algorithm, values converged to $q_0=0.7372$, $q_1=q_2=0.3960$, $R_0=0.1918$, which were obtained by exhaustive search in Fig.~\ref{fig:risk_contour}. We also obtained the same results (not shown) in Fig.~\ref{fig:opt_belief_trend}, which numerically verifies that having identical local beliefs does not lose optimality in the Gaussian setting. As Alg.~\ref{alg:PBPO} is vanilla and there are numerous optimization methods recently, further improvements are possible: For instance, one can use gradient-based or second-order methods as well as $\Delta$-scheduling methods. However, such algorithmic improvements are beyond our scope of this work.

Since the local agents' observations are i.i.d.~conditioned on $H$, the assumption of identical local beliefs is often made. This does not in general guarantee global optimality though, cf.~counterexamples can be found in \cite{TenneyS1981, Tsitsiklis1988} for standard distributed detection settings, it greatly simplifies analytic and numerical analysis. Note that the fixed point in Fig.~\ref{fig:opt_belief_trend}, i.e., $\pi_0 = q_\cent^* = q_1^* = \ldots =q_N^*$, is at $\frac{c_{\MD}}{ c_{\FA}+c_{\MD} }$. Restricting to identical local beliefs, we can analytically prove that it is a fixed point. Let us note a useful property of \eqref{eq:diff_zero} before stating it.

\begin{lem} \label{lem:strictly_dec}
	The right side of \eqref{eq:diff_zero} is strictly decreasing in $q_i$, $i \in \{1, \ldots, N\} \setminus \{j\}$, when other parameters are fixed.
\end{lem}
\begin{IEEEproof}
Provided in App.~\ref{app:strictly_inc}.
\end{IEEEproof}

Now the fixed point result follows.
\begin{cor} \label{thm:fixed_pt}
Suppose the noise is additive, independent, zero mean, and its distribution is symmetric, e.g., Gaussian noise. Then, among the set of identical beliefs, i.e., $q_1=\cdots=q_N$, knowing the true prior is optimal when $\pi_0 \in \{0, \frac{c_{\MD}}{ c_{\FA}+c_{\MD} }, 1 \}$.
\end{cor}
\begin{IEEEproof}
The cases $\pi_0 \in \{0,1\}$ are trivial so focus on $\pi_0 = \frac{c_{\MD}}{ c_{\FA}+c_{\MD} }$. At this $q_i^*$, each agent takes initial decision threshold $\lambda_i = 1/2$ by \eqref{eq:decision_threshold}. It implies by symmetry that
\begin{align*}
p_{\wh{H}_i|H}(1|0) = p_{\wh{H}_i|H}(0|1) \textrm{ and } p_{\wh{H}_i|H}(0|0) = p_{\wh{H}_i|H}(1|1).
\end{align*}
Furthermore, when the fusion agent's initial threshold is also $1/2$, it holds by symmetry that
\begin{align*}
p_{\wh{H}_\cent|H,\wh{H}_{-j}^N,\wh{H}_j}(1|0, \wh{h}_{-j}^N, h) = p_{\wh{H}_\cent|H,\wh{H}_{-j}^N,\wh{H}_j}(0|1, \neg \wh{h}_{-j}^N, \neg h),
\end{align*}
where $\neg (\cdot)$ stands for a flip of decision. Hence, $p_\FA(\wh{h}_j=1) =p_\MD(\wh{h}_j=0), p_\FA(\wh{h}_j=0) = p_\MD(\wh{h}_j=1)$, and therefore, \eqref{eq:diff_zero} hold. Since the right side of \eqref{eq:diff_zero} is decreasing along the $q_1 = \cdots = q_N$ direction, the solution is unique and optimal.
\end{IEEEproof}

\subsection{Human-AI Teams} \label{sec:human_ai}
\begin{figure}[t]
	\centering
	\includegraphics[width=3.1in]{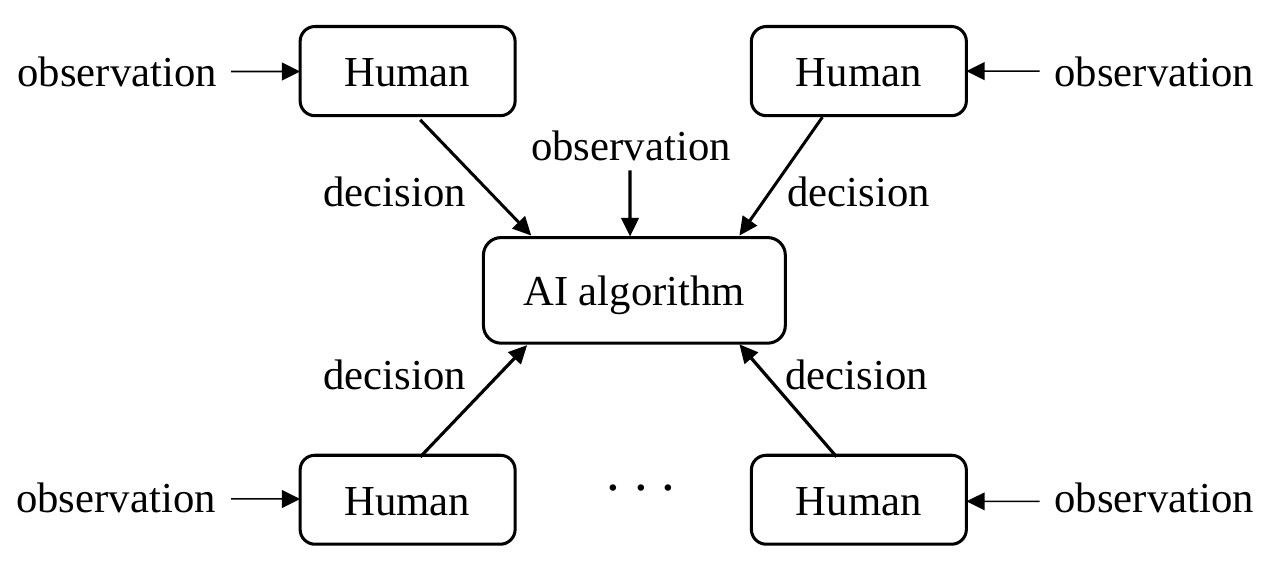}
	\caption{ A model of human-AI collaborative systems. Local human agents provide inputs for machine judgment.}
	\label{fig:human_AI}
\end{figure}
The model in this work assumes each agent perceives the prior differently from the truth. This phenomenon has been widely discussed in cumulative prospect theory, where people do not perceive probabilities linearly and tend to overweight small probabilities and underweight large probabilities \cite{TverskyK1992, GonzalezW1999}. Although agents in our model need not be humans, considering the local agents as humans could have important implications for socio-technical network inference systems.

Specifically, we study a problem of human-AI collaborative systems such as recommendation systems in e-commerce, depicted in Fig.~\ref{fig:human_AI}. To this end, we assume that human agents perform Bayesian decision making \cite{BraseCT1998, GlanzerHM2009}, and their perceptions follow the Prelec reweighting function \cite{Prelec1998} in Def.~\ref{def:prelec}. We also assume private observations are with additive standard Gaussian noise, common in human signal perception~\cite{KnillR1996, YostPF2003}. The fusion agent, possibly an AI algorithm, can be arbitrarily biased as we want, whereas human agents cannot be and only follow the Prelec weighting law. The result in Fig.~\ref{fig:opt_belief_trend} indicates that the optimal beliefs of the local agents (that is, human agents in this scenario) overweight small probabilities and underweight large probabilities. We approximate $q_1^*$ by the Prelec reweighting function in the minimax sense to see how humans agents perform for the decision-making system, that is, we find the Prelec parameters $\alpha^*, \beta^*$ such that 
\begin{align*}
(\alpha^*, \beta^*) = \argmin_{\alpha, \beta > 0} \|q_1^*(\cdot) - w(\cdot ; \alpha, \beta) \|_{\infty}.
\end{align*}

\begin{figure}[t]
	\centering
	\subfloat[$c_\FA=c_\MD=1$]{\includegraphics[width=2.3in]{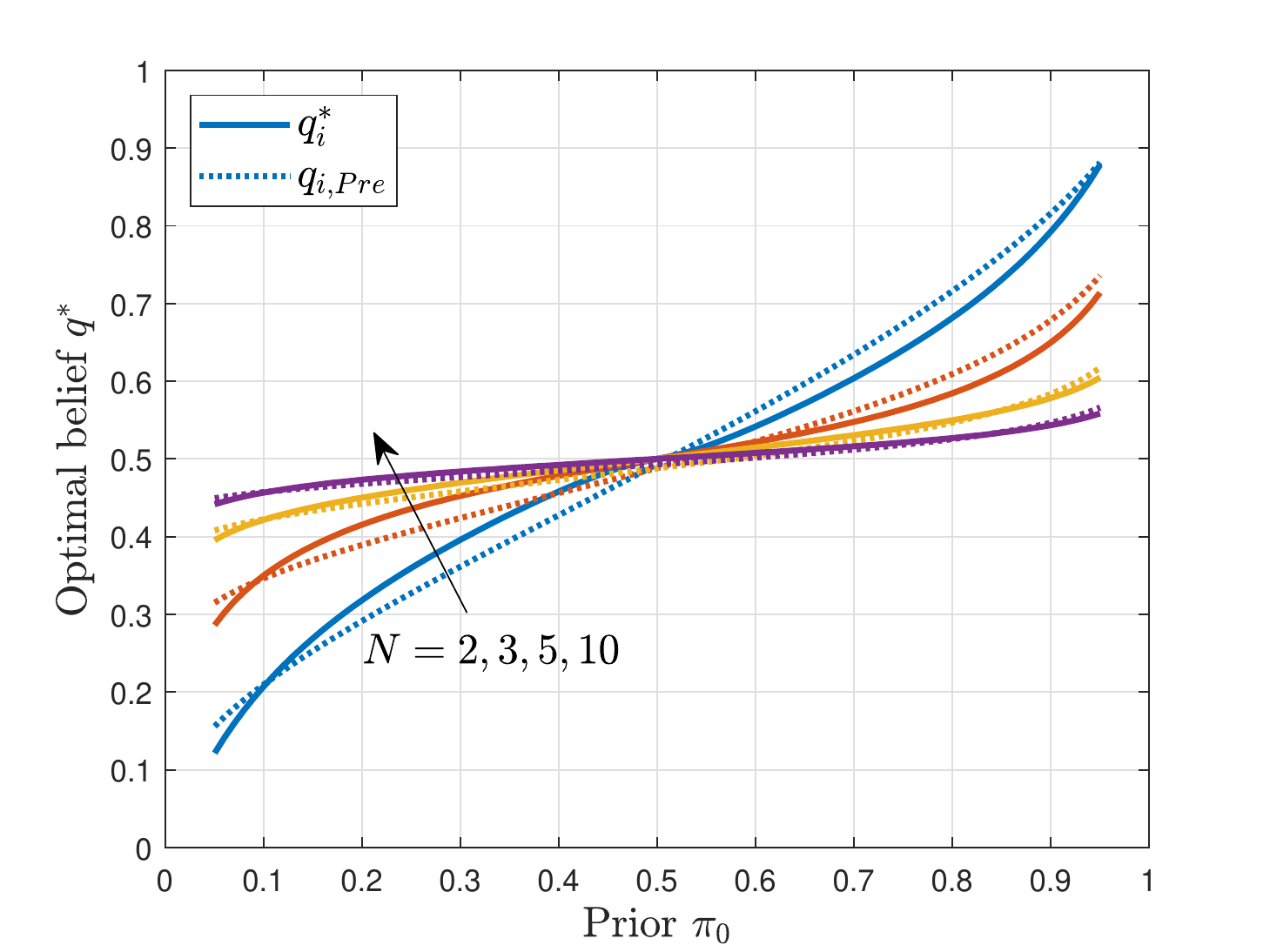}} \\	\subfloat[$c_\FA=1, c_\MD=2$]{\includegraphics[width=2.3in]{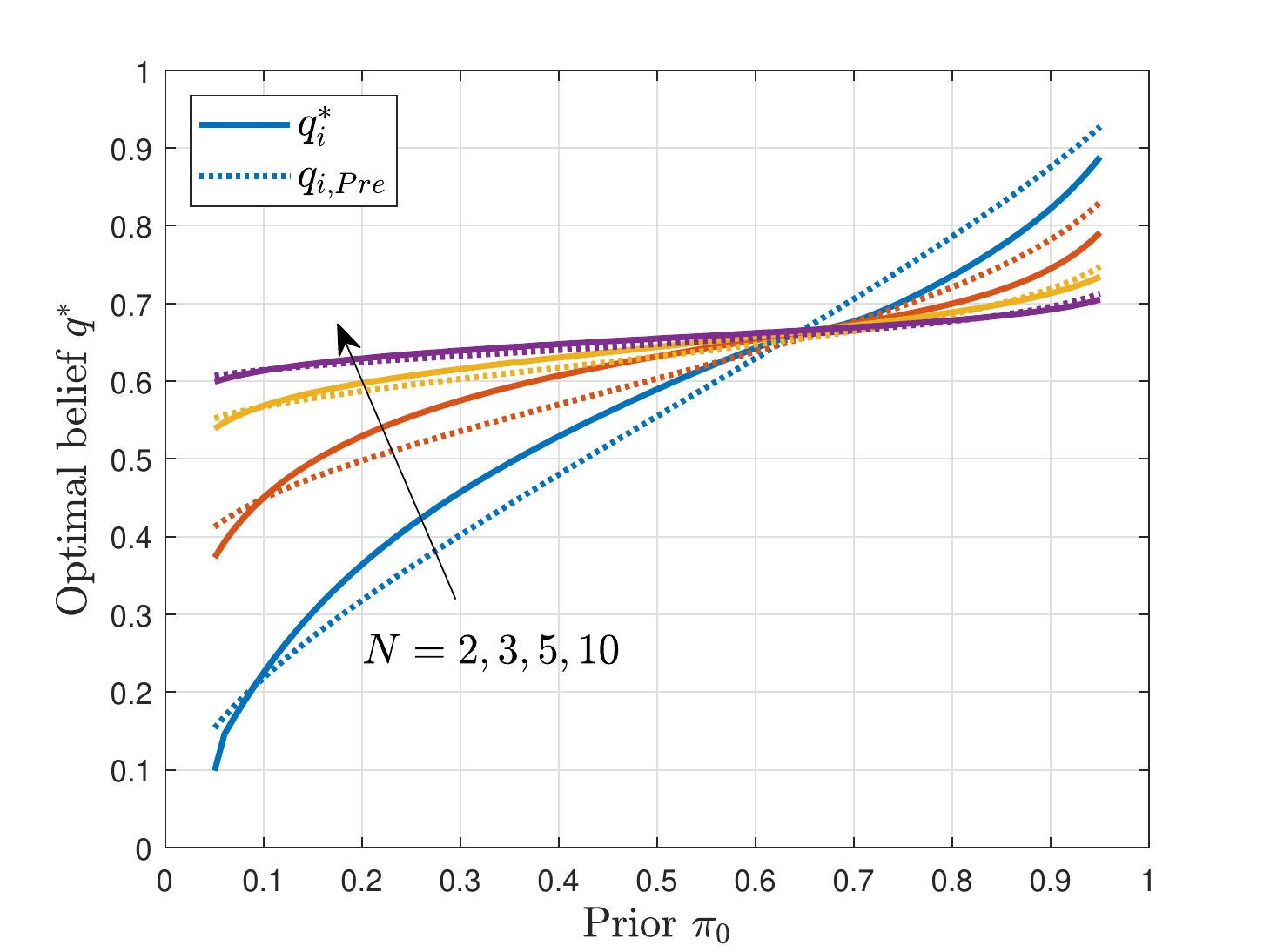}}	
	\caption{Optimal beliefs and their Prelec approximation. The noise is additive standard Gaussian. Solid curves are the optimal beliefs and dotted curves are  approximation by Prelec function. Arrows indicate ``as $N$ grows''.}
	\label{fig:prelec_curve}
\end{figure}
\begin{figure}[t]
	\centering
	\subfloat[$c_\FA=c_\MD=1$]{\includegraphics[width=1.82in]{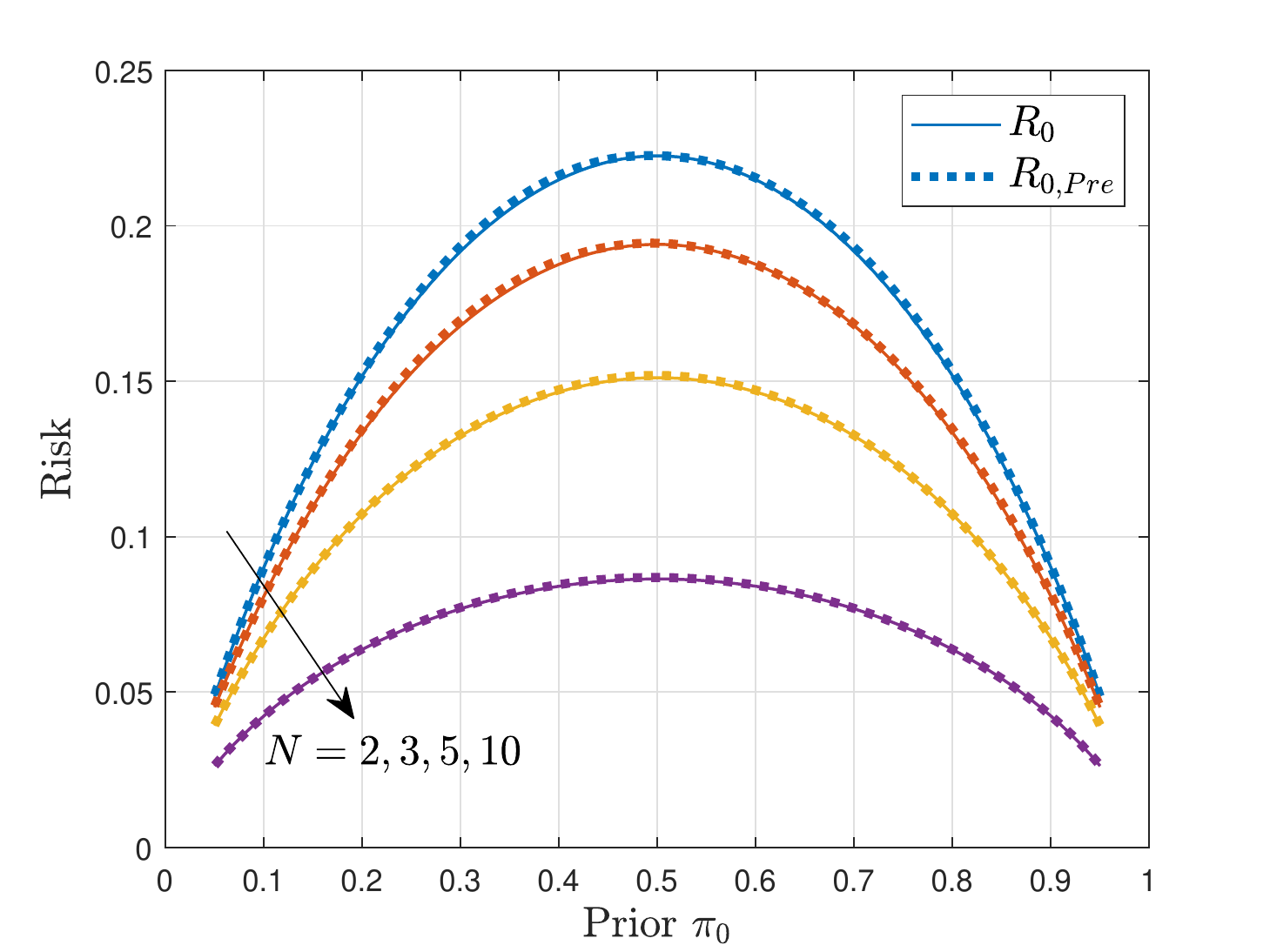}}
	\subfloat[$c_\FA=1, c_\MD=2$]{\includegraphics[width=1.82in]{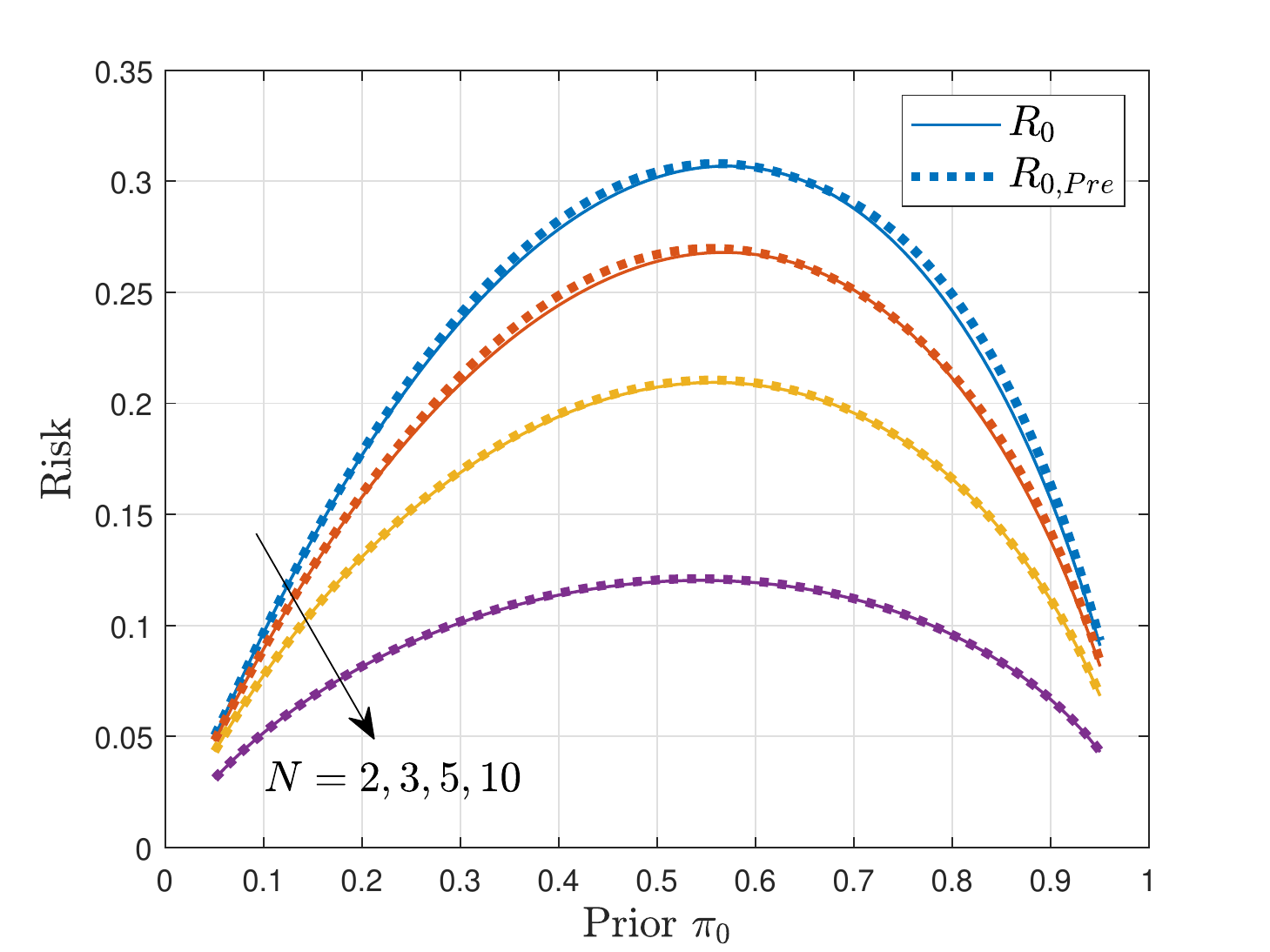}}
	\caption{Optimal risk and incurred risk by Prelec approximation. The noise is additive standard Gaussian. The optimal Bayes risks are in solid curve and incurred risks using Prelec beliefs are in dotted curve. Arrows indicate ``as $N$ grows''.}
	\label{fig:prelec_risk_curve}
\end{figure}

The approximation result is shown in Figs.~\ref{fig:prelec_curve} and \ref{fig:prelec_risk_curve}.\footnote{In numerical results, computations are only over $\pi_0 \in [0.05, 0.95]$ because of computational instability.} Surprisingly, the Prelec weighting function approximates $q_i^*$ well so that the extra loss due to the Prelec human agents is negligible. This indicates that the optimal belief conforms to some Prelec-weighted beliefs of humans, and so the irrationality of humans might be well-suited to group decision making in star networks. The largest extra loss of Bayes risk is, for instance of $N=2$ and equal costs, $\|R_{\cent, \textrm{Pre}} - R_0\|_{\infty} \approx 0.0015$ at $\pi_0=0.35$, while $R_\cent$ at $\pi_0=0.35$ is $0.2053$ so negligible. The close approximation suggests that a fusion agent holding a contrarian belief can work efficiently in the presence of human irrationality, facilitating improved inference in a near-optimal way.

Note that among many other reweighting functions, the Prelec function satisfies a set of axioms called {\em compound invariance} \cite{Prelec1998}, and therefore, people widely use the Prelec function for probability reweighting. The Prelec function is not cherry-picked for our approximation purpose since all reweighting functions have similar shapes, e.g., \cite{GonzalezW1999}, with only minor differences. Using another reweighting function does not significantly change our observation.

\section{Asymptotic Optimality of Identical Beliefs} \label{sec:asymp}
\begin{figure}[t]
	\centering
	\includegraphics[width=2.1in]{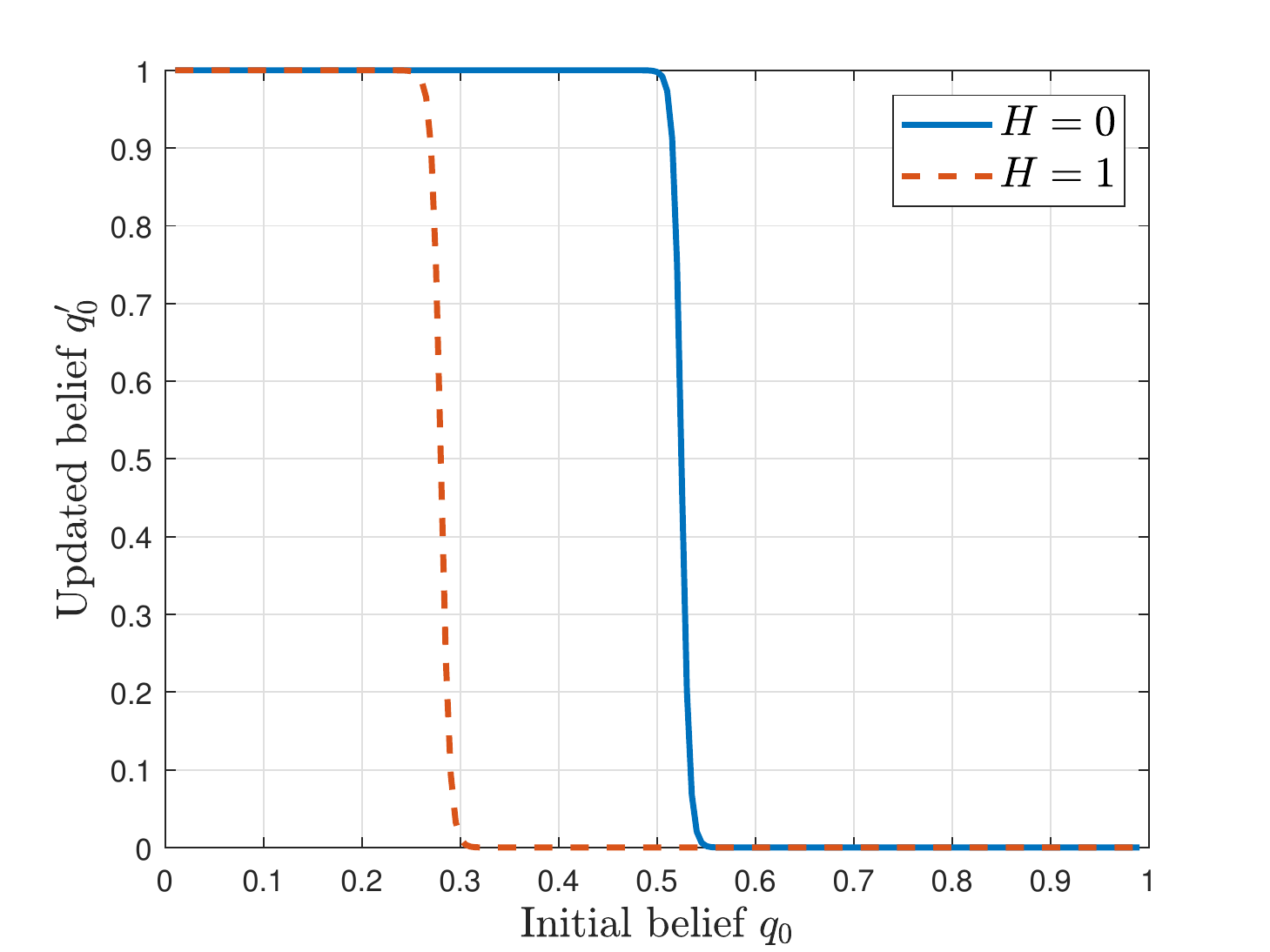}
	\caption{The phase transition of the updated belief \eqref{eq:conti_mapping} for $N=100$ local agents with $q_1 = \cdots = q_{100} = 0.4$. Additive standard Gaussian noise is assumed.}
	\label{fig:belief_polar}
\end{figure}

In this section, we consider asymptotics in many-agent star networks as $N \to \infty$. Our final goal is to derive the asymptotically optimal belief tuple $(q_\cent^*, q_1^*, \ldots, q_N^*)$ and its risk characterization. To this end, we begin with the assumption that local decision rules are all identical, i.e., $q_1 = \cdots = q_N$, but $q_\cent$ can be arbitrary. The choice of identical local beliefs often results in a suboptimal risk \cite{TenneyS1981, Tsitsiklis1988}, while greatly reduces the analytic and computational complexity. However, we will see that the loss due to the identical beliefs is provably asymptotically negligible. Recall the results for finite $N$ that optimal beliefs $q_\cent^*, q_i^*$ are dependent on $\pi_0$ so the external system designer must have knowledge of the true prior to attain the least Bayes risk. However, unlike finite $N$, we now show that the choice of $q_\cent = q_1 = \cdots = q_N = \text{constant}$ regardless of $\pi_0$ is asymptotically optimal, as in Thm.~\ref{thm:asymp_optimal_belief}.

Let us first assume $q_1 = \cdots = q_N$ and observe the properties of agents' behavior. As the private observations are i.i.d.~and depend on $H$, and each local agent decides as if $q_1$ is the true prior, local decisions depend on $H$ and are i.i.d.~as well. To be precise, let $\mathcal{A}_1$ be the local agent's decision region for $\wh{H}_i=1$, i.e., $\mathcal{A}_1$ is the collection of $y$ such that
\begin{align*}
	\frac{f_{Y|H}(y_i|1)}{f_{Y|H}(y_i|0)} > \frac{c_\FA q_1}{c_\MD (1- q_1)}.
\end{align*}
Then, false alarm and missed detection probabilities can be written as:
\begin{align*}
	p_{\wh{H}_i |H} (1|0) &= \int_{\mathcal{A}_1}  f_{Y|H}(dy|0), \\
	p_{\wh{H}_i |H} (0|1) &= \int_{\mathcal{A}_1^c} f_{Y|H}(dy|1),
\end{align*}
which are deterministic functions of $q_1$. According to the property of the receiver operating characteristic (ROC) \cite{Poor1988}, $t_0 \triangleq p_{\wh{H}_i |H} (1|0) < t_1 \triangleq p_{\wh{H}_i |H} (1|1)$ for all $q_1$, that is, $\wh{H}_i=1$ is more likely when $H=1$.

The next proposition discusses the phase transition of the updated belief when $N$ is large, which in turn implies the decision is close to a \emph{deterministic} function in $(q_0, q_1)$. It is illustrated in Fig.~\ref{fig:belief_polar}: As the updated belief is close to either $0$ or $1$, the decision is close to being deterministic.
\begin{prop} \label{thm:polarization}
When $q_1 = \cdots = q_N$, the updated belief \eqref{eq:belief_update} of the fusion agent approaches either $0$ or $1$ almost surely as $N \to \infty$. The limit point is a deterministic function of $(q_\cent, q_1)$.
\end{prop}
\begin{IEEEproof}
Consider the belief update formula \eqref{eq:belief_update} for $(\wh{h}_1, \ldots, \wh{h}_N)$ and define a random variable $r_1$ to be the ratio of decision $1$'s in $\wh{h}^N$, i.e., $r_1 \triangleq \frac{\# \textrm{ of ones in } \wh{h}^N}{N}$. For brevity let
\begin{align*}
	z_1 \triangleq \frac{p_{\wh{H}_i|H}(0|0)_{[\cent]}}{p_{\wh{H}_i|H}(0|1)_{[\cent]} }, z_2 \triangleq \frac{p_{\wh{H}_i|H}(0|1)_{[\cent]}}{p_{\wh{H}_i|H}(0|0)_{[\cent]} } \cdot \frac{p_{\wh{H}_i|H}(1|0)_{[\cent]}}{p_{\wh{H}_i|H}(1|1)_{[\cent]} }.
\end{align*}

Let $q_0'$ be the updated prior at the fusion agent. Then, algebraic manipulation shows the following.
\begin{align}
	&\frac{q_\cent'}{1-q_\cent'} = \frac{q_\cent}{1-q_\cent} \prod_{i=1}^N \frac{p_{\wh{H}_i|H}(\wh{h}_i|0)_{[\cent]}}{p_{\wh{H}_i|H}(\wh{h}_i|1)_{[\cent]} } \nonumber \\
	&= \frac{q_\cent}{1-q_\cent} \left( \frac{p_{\wh{H}_i|H}(0|0)_{[\cent]}}{p_{\wh{H}_i|H}(0|1)_{[\cent]} } \right)^{\textrm{\# of $0$s}} \left( \frac{p_{\wh{H}_i|H}(1|0)_{[\cent]}}{p_{\wh{H}_i|H}(1|1)_{[\cent]} } \right)^{\textrm{\# of $1$s}} \nonumber \\
	&= \frac{q_\cent}{1-q_\cent} \left[ \left( \frac{p_{\wh{H}_i|H}(0|0)_{[\cent]}}{p_{\wh{H}_i|H}(0|1)_{[\cent]} } \right)^{1-r_1} \left( \frac{p_{\wh{H}_i|H}(1|0)_{[\cent]}}{p_{\wh{H}_i|H}(1|1)_{[\cent]} } \right)^{r_1} \right]^N \nonumber \\
	&= \frac{q_\cent}{1-q_\cent} \left( z_1(q_0) z_2(q_0)^{r_1(q_1)} \right)^N, \label{eq:def_of_alpha}
\end{align}
In addition, $\wh{h}_1, \ldots, \wh{h}_N$ are $N$ i.i.d.~copies of Bernoulli random variable with $\mathbb{P}[\wh{h}_i=1|H=0] = t_0(q_1)$ if $H=0$, and $\mathbb{P}[\wh{h}_i=1|H=1] = t_1(q_1)$ if $H=1$. In other words,
\begin{align}
	r_1 \to
	c_H \triangleq \begin{cases}
	t_0(q_1) & \textrm{if } H=0, \\
	t_1(q_1) & \textrm{if } H=1,
	\end{cases} \label{eq:r0_asymp}
\end{align}
almost surely as $N$ grows. As $r_1$ converges to $c_H > 0$,
\begin{align}
	& \lim_{N \to \infty} \frac{q_\cent}{1-q_\cent} \left( z_1 z_2^{r_1} \right)^N = \lim_{N \to \infty} \frac{q_\cent}{1-q_\cent} \left( z_1 z_2^{c_H} \right)^N \nonumber \\
	&= \begin{cases}
	\lim_{N \to \infty} \frac{q_\cent}{1-q_\cent} \left( z_1 z_2^{t_0} \right)^N & \textrm{if } H=0, \\
	\lim_{N \to \infty} \frac{q_\cent}{1-q_\cent} \left( z_1 z_2^{t_1} \right)^N & \textrm{if } H=1.
	\end{cases} \label{eq:conti_mapping}
\end{align}
Depending on the value to be exponentiated, the right side approaches either $0$ or $\infty$. Since $x/(1-x):(0,1) \mapsto (0, \infty)$ is monotonic in $x \in (0,1)$, the updated belief shows phase transition.
\end{IEEEproof}

\begin{table}[b]
\center
\caption{Asymptotic risk of the fusion agent as a function of initial beliefs.} \label{tab:partition}
\begin{tabular}[t]{c | c | c || c }
	\hline
	& $z_1 z_2^{t_0}$ & $z_1 z_2^{t_1}$ & Limit value of $R_\cent$ \\
	\hline\hline
	\textsc{case} 1 & $>1$ & $<1$ & $0$ \\ 
	\hline
	\textsc{case} 2 & $<1$ & $<1$ & $c_{\FA} \pi_0$ \\
	\hline
	\textsc{case} 3 & $>1$ & $>1$ & $c_{\MD} \bar{\pi}_0$ \\
	\hline
	\textsc{case} 4 & $<1$ & $>1$ & impossible \\
	\hline
\end{tabular}
\end{table}
\begin{figure}[t]
	\centering
	\includegraphics[width=2.4in]{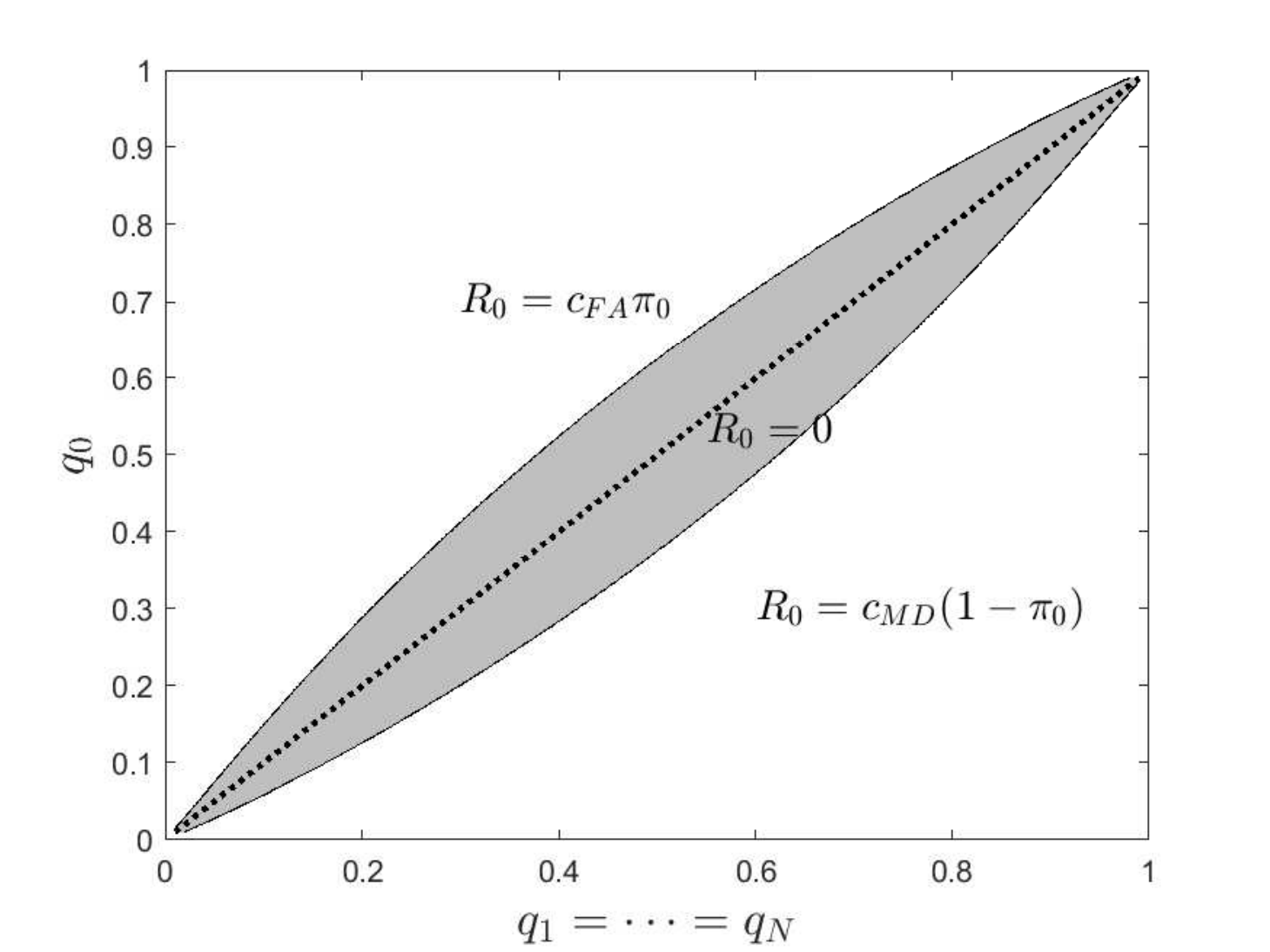}
	\caption{Beliefs partition by limiting value of $R_\cent \in \{0, c_{\FA} \pi_0, c_{\MD}\bar{\pi}_0\}$. The optimal points for large $N$ suggested by Fig.~\ref{fig:opt_belief_trend}, i.e., $\left(\frac{c_{\MD}}{ c_{\FA}+c_{\MD} }, \ldots, \frac{c_{\MD}}{ c_{\FA}+c_{\MD} }, \frac{c_{\MD}}{ c_{\FA}+c_{\MD} }\right)$, are drawn in dotted line. The noise is additive standard Gaussian.}
	\label{fig:risk_partition}
\end{figure}

Prop.~\ref{thm:polarization} reveals the fact that when $N$ is large, the fusion agent makes a decision that is asymptotically deterministic, as a function of $q_1$ and $q_\cent$ no matter what value the private signal $Y_\cent$ takes. Also, note that the true prior $\pi_0$ does not affect the decision by the fusion agent. It only affects $R_\cent$. This will be a crucial observation in the proof of Thm.~\ref{thm:asymp_optimal_belief}.

Updating the belief, the fusion agent always makes a correct decision if $q_\cent' = 1$ when $H=0$ and $q_\cent'=0$ when $H=1$. Therefore, due to the property $t_0 \triangleq p_{\wh{H}_i |H} (1|0) < t_1 \triangleq p_{\wh{H}_i |H} (1|1)$, the fusion agent is always asymptotically correct at least for one hypothesis, no matter what $(\pi_0, q_\cent, q_1)$ tuple is used. As summarized in Tab.~\ref{tab:partition}, the case of always being wrong (i.e., \textsc{case} $4$) is impossible. We include the proof for this statement in App.~\ref{app:impossible} for completeness.

Among the possible three cases, the shaded region in Fig.~\ref{fig:risk_partition} illustrates $R_\cent \to 0$ case for additive standard Gaussian noise. Two important observation can be made: First, the shaded region contains $\frac{c_{\MD}}{ c_{\FA}+c_{\MD} }=q_\cent = q_1=\cdots = q_N$ (the dotted line) for any $c_{\FA}, c_{\MD}$, at which $R_\cent$ asymptotically vanishes regardless of $\pi_0$ as suggested by Fig.~\ref{fig:opt_belief_trend}. Second, for any given $q_1$, the fusion agent can obtain an arbitrarily small risk by taking an appropriate $q_0$ so that $(q_0, q_1)$ belongs to the shaded region. Consequently, taking an appropriate $q_0$ according to $q_1$ is important. Note that setting the appropriate belief can be done by announcing an incorrect prior on purpose, e.g., an external organizer of the system misinforms the fusion agent so that it perceives as an appropriate $q_1$. Furthermore, the risk converges to its limit value at worst exponentially fast as rigorously proven in Prop.~\ref{thm:risk_exponent} and Thm.~\ref{thm:asymp_optimal_belief}.

We can also derive the speed of risk convergence to its limiting value in Fig.~\ref{fig:risk_partition}. To explicitly denote dependency on $N$, let $R_\cent^{(N)}$ be the risk of the fusion agent with $N$ local agents and $R_\cent^{(\infty)} \triangleq \lim_{N \to \infty} R_\cent^{(N)} \in \{0, c_{\FA}\pi_0, c_{\MD}\bar{\pi}_0\}$. Then, the next theorem shows that $R_\cent^{(N)} \to R_\cent^{(\infty)}$ exponentially fast in $N$, that is,
\begin{align*}
\beta \triangleq \sup \left( - \lim_{N \to \infty} \frac{1}{N} \log \left( R_\cent^{(N)} - R_\cent^{(\infty)} \right) \right) > 0,
\end{align*}
where the supremum is over all decision rules \emph{described in Sec.~\ref{subsec:belief_update}}.

\begin{prop} \label{thm:risk_exponent}
Suppose the private observataions are sub-Gaussian. Also, suppose that $(q_\cent, q_1)$ satisfies \textsc{case} $1$, $2$, or $3$, that is, $(q_\cent, q_1)$ strictly belongs to one of the regions in Fig.~\ref{fig:risk_partition}. Then, $\beta$ is strictly positive.
\end{prop}
\begin{IEEEproof}
Proof is based on the concentration inequality of i.i.d.~Bernoulli random variables and sub-Gaussian property of distribution tails. Details are in App.~\ref{app:pf_of_risk_exp}.
\end{IEEEproof}

Returning to the result for finite $N$ in Fig.~\ref{fig:opt_belief_trend}, two important observations can be made. The first is that we cannot achieve the smallest Bayes risk if the external network organizer does not know $\pi_0$ since the optimal beliefs are functions of $\pi_0$. The other is that the optimal beliefs converge to a constant ($\frac{c_\MD}{c_\FA + c_\MD}$ for Gaussian noise) as $N$ grows, although curves for $N=5, 10$ in Fig.~\ref{fig:opt_belief_trend} are drawn under $q_1 = \cdots = q_N$ assumption. We will see that the first observation is no longer true when $N \to \infty$, and the second one is analytically provable.

Denote the optimal risk exponent by
\begin{align*}
	\beta^* &\triangleq \sup \left( - \lim_{N \to \infty} \frac{1}{N} \log R_\cent^{(N)} \right),
\end{align*}
where the supremum is over all decision rules, not limited to identical rules or LRTs given in Sec.~\ref{subsec:belief_update}, but \emph{including strategic decision rules}, e.g., \cite{Varshney1997, Tsitsiklis1988}. Unlike observations for finite $N$, the next theorem states that $\beta^*$ can be attained simply by setting $q_\cent = q_1 =\cdots =q_N=\frac{c_\MD}{c_\FA + c_\MD}$ for any $\pi_0$ under some assumption.

\begin{thm} \label{thm:asymp_optimal_belief}
	Suppose the private observations are sub-Gaussian. Then, there exists some optimal identical beliefs at all agents $q^* = q_\cent = q_1 = \cdots = q_N$ and performing the LRT with belief update \eqref{eq:belief_update} asymptotically achieves $\beta^*$ as $N \to \infty$. Further, this achieves the optimal risk exponent of cooperative distributed detection in star networks in \cite{Tsitsiklis1988}, which is,
	\begin{align*}
		\beta^* = \max_{q_1} C(\textsf{Bern}(p_{\wh{H}_1|H}(1|0)), \textsf{Bern}(p_{\wh{H}_1|H}(1|1)) ).
	\end{align*}
\end{thm}
\begin{IEEEproof}
	Tsitsiklis \cite{Tsitsiklis1988} showed that an identical local decision rule is asymptotically optimal for the cooperative setting where the fusion agent knows all local decision rules. It also differs from ours in that error probability (i.e., unit costs) and no private observation at the fusion agent (i.e., $Y_\cent = \emptyset$) are considered. Building on the results so far and adapting \cite{Tsitsiklis1988}, we can show the claims. Details are in App.~\ref{app:pf_by_tsitsiklis}.
\end{IEEEproof}
The theorem indicates that the performance loss due to selfishness and two mismatches (between the true prior and beliefs, and between the $i$-th agent and $j$-th agent's beliefs) can be made insignificant by taking good identical beliefs. In addition, the optimal belief indeed equalizes the false alarm and missed detection error probabilities---this corroborates our knowledge that when there are many observations, simply equalizing the error exponents is asymptotically optimal, i.e., the maximum a posteriori (MAP) estimation and maximum likelihood (ML) estimation make no significant difference in asymptotics \cite{MoulinV2019}.

Like the finite network case, a clever organizer of the decision-making system would provide incorrect information about the true prior to minimize the risk of the fusion agent. To do so, even organizers of the many-agent networks need not know the true prior! Simply revealing incorrect information (or revealing biased training data if considering labeling by crowdworkers) so that the local agents perceive $q^*$ as the true prior is asymptotically optimal.

From the statement of Thm.~\ref{thm:asymp_optimal_belief}, it is immediate that $\beta^*$ is independent of $\pi_0, c_\FA, c_\MD$ and the presence of $Y_0$. Independence from $\pi_\cent, c_\FA, c_\MD$ is essentially because $R_\cent$ is dominated by the exponential decay of false alarm and missed detection error probabilities as $N \to \infty$. The presence of $Y_\cent$ is also asymptotically negligible under the assumption of sub-Gaussianity since local decisions dominate as $N$ grows. This can be also seen from the arguments that local decisions outweigh the perceived belief at the end of Sec.~\ref{subsec:belief_update} and that the fusion agent's decision is asymptotically deterministic in $q_\cent, q_1$.

Finally, we can obtain the optimal belief and risk exponent explicitly if the noise is additive Gaussian.
\begin{cor} \label{cor:Gaussian_risk_exponent}
	Suppose additive Gaussian noise. Then, having $q_\cent = q_1 = \cdots = q_N = \frac{c_\MD}{c_\FA + c_\MD}$ belief minimizes the Bayes risk $R_\cent$ asymptotically. Furthermore, 
	\begin{align*}
		\beta^* &= C(\textsf{Bern}(Q(0.5/\sigma)), \textsf{Bern}(Q(-0.5/\sigma))) \\
		&\approx 0.0793 ~~~ \textrm{if $\sigma = 1$.}
	\end{align*}
	where $C(\cdot, \cdot)$ is the Chernoff information.
\end{cor}
\begin{IEEEproof}
	Given in App.~\ref{app:pf_gaussian}.
\end{IEEEproof}

\section{Conclusion and Discussion} \label{sec:discussion}
This work investigates a decision-making problem in star networks with two distinguishing aspects that 1) agents are selfish, and 2) agents perceive the prior differently, and further the fusion agent thinks all local agents believe as it believes. In this setting, we show that the belief tuple that minimizes the fusion agent's Bayes risk is in general different from the tuple of true priors. The optimal local agent should overweight small prior probabilities and underweight large ones, so it has a distorted view of the hypothesis. However, the optimal fusion agent should have a contrary view against the true prior and the local agents. Also, the optimal belief of the local agents can be accurately approximated by a human agent following cumulative prospect theory. Therefore under our model, decisions made by people are nearly optimal for a machine to make a final decision. The setting where the number of local agents is large is also investigated. As suggested from the numerical result for a finite number of agents, it turns out that the optimal beliefs are asymptotically identical no matter what the true prior is. Also, the optimal risk asymptotically vanishes exponentially fast and the exponent is explicitly characterized.

This work opens up several future directions as inference over networks of selfish agents having biased perception is a new model in distributed detection. An immediate issue is to extend the current results to $M \ge 3$ hypothesis cases. Recall that the standard optimal decision rule in $M$-ary hypothesis testing consists of $(M-1)$ binary likelihood ratio tests and only one test is active for a particular observation. This incurs a discontinuity in each decision threshold, so we need a new approach instead of differentiation in the proof of Thm.~\ref{thm:derivative_zero}. For asymptotics, Tsitsiklis \cite{Tsitsiklis1988} considered a cooperative star network where the fusion agent knows the decision rules that local agents used. Then, for $M \ge 3$, it is shown that dividing $N$ local agents into $M(M-1)/2$ groups with each group having identical decision rules is asymptotically optimal. Hence, it is also worth seeing whether $M(M-1)/2$ identical rules still hold in our setting.

Another promising direction is to investigate the case with more general noise. Sub-Gaussianity in Sec.~\ref{sec:asymp} is perhaps an easy extension of Gaussian noise and limits the effect of tails of private observations. Therefore, it is of interest to see how our results should be changed for heavy-tailed distributions.

\section*{Acknowledgment}
The authors thank an anonymous reviewer of \cite{SeoRRGV2019} for suggesting this investigation, and J.~B.~Rhim and V.~K.~Goyal for helpful discussions. The authors also thank the anonymous reviewers for their comments greatly improving our work.

\appendices
\section{Proof of Thm.~\ref{thm:derivative_zero}} \label{app:derivative_zero}
Without loss of generality, suppose that at the decision threshold $\lambda_j$ of our interest, increasing $\lambda_j$ decreases the false alarm probability, but increases the missed detection probability. Otherwise, the same proof still applies with opposite signs.

Note that from \eqref{eq:q_function}, for $j \in \{1,\ldots,N\}$,
\begin{subequations} \label{eq:error_derivative}
\begin{align}
	\frac{\partial p_{\wh{H}_j|H}(1|0)}{\partial \lambda_j} &= -f_{Y|H}(\lambda_j|0), \\
	\frac{\partial p_{\wh{H}_j|H}(0|1)}{\partial \lambda_j} &= f_{Y|H}(\lambda_j|1).
\end{align}
\end{subequations}

Differentiating \eqref{eq:R_N} with respect to $\lambda_j, j\in\{1,\ldots,N\}$ using \eqref{eq:error_derivative}, we have \eqref{eq:long_eq} in the next page.
\begin{table*}[h]
	\begin{equation} \label{eq:long_eq}
	\begin{aligned}
	\frac{\partial R_\cent}{\partial \lambda_j} &= c_{\FA} \pi_0 f_{Y|H}(\lambda_j|0) \Bigg[ \underbrace{ \sum_{\wh{h}_{-j}^N} \Bigg( \prod_{i \neq j} p_{\wh{H}_i|H}(\wh{h}_i|0) \Bigg) p_{\wh{H}_\cent|H,\wh{H}^N}(1|0, \wh{h}_{-j}^N, 0) }_{= p_\FA(\wh{h}_j=0)} - \underbrace{ \sum_{\wh{h}_{-j}^N} \Bigg( \prod_{i \neq j} p_{\wh{H}_i|H}(\wh{h}_i|0) \Bigg) p_{\wh{H}_\cent|H,\wh{H}^N}(1|0, \wh{h}_{-j}^N, 1)}_{= p_\FA(\wh{h}_j=1)} \Bigg] \\
	&+ c_{\MD}\bar{\pi}_0 f_{Y|H}(\lambda_j|1) \Bigg[ \underbrace{ \sum_{\wh{h}_{-j}^N} \Bigg( \prod_{i \neq j} p_{\wh{H}_i|H}(\wh{h}_i|1) \Bigg) p_{\wh{H}_\cent|H,\wh{H}^N}(0|1, \wh{h}_{-j}^N, 0)}_{= p_\MD(\wh{h}_j=0)} - \underbrace{ \sum_{\wh{h}_{-j}^N} \Bigg( \prod_{i \neq j} p_{\wh{H}_i|H}(\wh{h}_i|1) \Bigg) p_{\wh{H}_\cent|H,\wh{H}^N}(0|1, \wh{h}_{-j}^N, 1)}_{= p_\MD(\wh{h}_j=1)} \Bigg].
	\end{aligned}
	\end{equation}
	\hrule
\end{table*}

Setting the derivative zero and rearranging terms,
\begin{align*}
\frac{f_{Y|H}(\lambda_j|1)}{f_{Y|H}(\lambda_j|0)} = \frac{c_{\FA}\pi_0}{c_{\MD}\bar{\pi}_0} \frac{p_\FA(\wh{h}_j=1) - p_\FA(\wh{h}_j=0)}{p_\MD(\wh{h}_j=0) - p_\MD(\wh{h}_j=1)}.
\end{align*}
Furthermore, we know from \eqref{eq:likelihood_test_distributed} that the left side is $\frac{c_{\FA}q_j^*}{c_{\MD}(1-q_j^*)}$. Therefore the claim has been proved that
\begin{align*}
\frac{q_j^*}{1-q_j^*} = \frac{\pi_0}{1-\pi_0} \frac{p_\FA(\wh{h}_j=1) - p_\FA(\wh{h}_j=0)}{p_\MD(\wh{h}_j=0) - p_\MD(\wh{h}_j=1)}.
\end{align*}

\section{Proof of Lem.~\ref{lem:convexity}} \label{app:convexity}
Let us focus on the agent $j \ne 0$. Rearranging \eqref{eq:risk_expression} in terms of $p_{\wh{H}_j|H}(1|0)$ and $p_{\wh{H}_j|H}(1|1)$ using $p_\FA(\wh{h}_j=h)$ and $p_\MD(\wh{h}_j=h)$ defined in \eqref{eq:def_p_FA_p_MD} gives
\begin{align*}
R_\cent &= c_{\FA} \pi_0 p_{\wh{H}_j|H}(0|0) p_\FA(\wh{h}_j=0) \\
&\quad \quad + c_{\FA} \pi_0 p_{\wh{H}_j|H}(1|0) p_\FA(\wh{h}_j=1) \\
&\quad \quad + c_{\MD}\bar{\pi}_0 p_{\wh{H}_j|H}(0|1) p_\MD(\wh{h}_j=0) \\
&\quad \quad + c_{\MD}\bar{\pi}_0 p_{\wh{H}_j|H}(1|1) p_\MD(\wh{h}_j=1) \\
&= c_{\FA} \pi_0 (1-p_{\wh{H}_j|H}(1|0)) p_\FA(\wh{h}_j=0) \\
&\quad \quad + c_{\FA} \pi_0 p_{\wh{H}_j|H}(1|0) p_\FA(\wh{h}_j=1) \\
&\quad \quad + c_{\MD}\bar{\pi}_0 (1-p_{\wh{H}_j|H}(1|1)) p_\MD(\wh{h}_j=0) \\
&\quad \quad + c_{\MD}\bar{\pi}_0 p_{\wh{H}_j|H}(1|1) p_\MD(\wh{h}_j=1) \\
&= c_{\FA} \pi_0 p_{\wh{H}_j|H}(1|0) \left(p_\FA(\wh{h}_j=1) - p_\FA(\wh{h}_j=0)\right) \\
&\quad \quad - c_{\MD} \bar{\pi}_0 p_{\wh{H}_j|H}(1|1) \left(p_\MD(\wh{h}_j=0) - p_\MD(\wh{h}_j=1)\right) \\
&\quad \quad + c_{\FA} \pi_0 p_\FA(\wh{h}_j=0) + c_{\MD}\bar{\pi}_0 p_\MD(\wh{h}_j=0).
\end{align*}

Now recall what $p_\FA(\wh{h}_j=h), p_\MD(\wh{h}_j=h)$ stand for---$p_\FA(\wh{h}_j=h)$ (or $p_\MD(\wh{h}_j=h)$) is the false alarm (or missed detection) probability of the fusion agent conditioned on $\wh{h}_j=h$. Also, recall that conditioning on $\wh{h}_j=0$ increases the fusion agent's initial belief, which in turn implies the decision threshold also does, whereas conditioning on $\wh{h}_j=1$ decreases the decision threshold. Since the false alarm probability is decreasing in the decision threshold, we can conclude that $p_\FA(\wh{h}_j=1)-p_\FA(\wh{h}_j=0)$ is nonnegative always. A similar argument for missed detection shows $p_\MD(\wh{h}_j=0) - p_\MD(\wh{h}_j=1)$ is nonnegative. Finally the fact from the property of a receiver operating characteristic curve \cite{Poor1988} that $p_{\wh{H}_j|H}(1|1)$ is strictly concave in $p_{\wh{H}_j|H}(1|0)$ yields the convexity in $p_{\wh{H}_j|H}(1|0)$.

For $p_{\wh{H}_\cent|H}(1|0)$, it can be shown similarly, i.e., $p_{\wh{H}_\cent|H}(1|1)$ is strictly concave in $p_{\wh{H}_\cent|H}(1|0)$ when $\{p_{\wh{H}_j|H}(1|0)\}_{j \ge 1}$ are fixed, or equivalently, $\{q_j\}_{j \ge 1}$ are fixed. Hence, \eqref{eq:R_N} is concave in $p_{\wh{H}_\cent|H}(1|0)$ as well.

\section{Proof of Lem.~\ref{lem:strictly_dec}} \label{app:strictly_inc}
In the proof of Lem.~\ref{lem:convexity}, it has been shown that $p_\FA(\wh{h}_j=1) - p_\FA(\wh{h}_j=0)$ and $p_\MD(\wh{h}_j=0) - p_\MD(\wh{h}_j=1)$ are positive.

Next investigate how the decision of agent $i \neq j$ changes $p_\FA(\wh{h}_j=1) - p_\FA(\wh{h}_j=0)$. Note that $p_\FA(\wh{h}_j=1) - p_\FA(\wh{h}_j=0)$ is linear in $p_{\wh{H}_i|H}(1|0) \in [0,1]$, i.e., the false alarm probability of agent $i$. To determine whether the slope is positive or negative, check two extremes $p_{\wh{H}_i|H}(1|0)=0$ and $p_{\wh{H}_i|H}(1|0)=1$. When $p_{\wh{H}_i|H}(1|0)=0$, in other words, $\wh{h}_i=0$ always, $p_\FA(\wh{h}_j=1) - p_\FA(\wh{h}_j=0)$ gives the false alarm probability conditioned on $\wh{h}_i=0$. Similarly $p_{\wh{H}_i|H}(1|0)=1$ corresponds to the false alarm probability conditioned on $\wh{h}_i=1$. The fact that observing decision $0$ increases the belief $q_\cent$ (hence $\lambda_\cent$ as well) implies $p_\FA(\wh{h}_j=1) - p_\FA(\wh{h}_j=0)$ at $p_{\wh{H}_i|H}(1|0)=0$ is smaller, which in turn implies that the slope is positive, or equivalently $p_\FA(\wh{h}_j=1) - p_\FA(\wh{h}_j=0)$ is decreasing in $q_i$ since $p_{\wh{H}_i|H}(1|0)$ is decreasing in $q_i$. Repeating the argument for $p_\MD(\wh{h}_j=h)$, we can conclude that $p_\MD(\wh{h}_j=0) - p_\MD(\wh{h}_j=1)$ is increasing in $q_i$, so the overall function
\begin{align*}
\frac{p_\FA(\wh{h}_j=1) - p_\FA(\wh{h}_j=0)}{p_\MD(\wh{h}_j=0) - p_\MD(\wh{h}_j=1)}
\end{align*}
is decreasing in $q_i$.

\section{\textsc{case} $4$ is Impossible} \label{app:impossible}
\begin{prop}
	The case when the fusion agent is always wrong for $H=0$ and $H=1$ both is impossible.
\end{prop}
\begin{IEEEproof}
	From the property of a receiver operating characteristic curve \cite{Poor1988}, $t_0(q) \triangleq p_{\wh{H}_i |H} (1|0) < t_1(q) \triangleq p_{\wh{H}_i |H} (1|1)$ always. Let us consider $z_2(q_0)$ term. Using the definition of $t_i(q)$, it can be rewritten as follows.
	\begin{align*}
		z_2 &= \frac{p_{\wh{H}_i|H}(0|1)_{[\cent]}}{p_{\wh{H}_i|H}(0|0)_{[\cent]} } \cdot \frac{p_{\wh{H}_i|H}(1|0)_{[\cent]}}{p_{\wh{H}_i|H}(1|1)_{[\cent]} } \\
		&= \frac{1-t_1(q_0)}{1-t_0(q_0)} \cdot \frac{t_0(q_0)}{t_1(q_0)} \\
		&\stackrel{(a)}{<} \frac{1-t_1(q_0)}{1-t_0(q_0)} \cdot 1 \stackrel{(b)}{<} 1,
	\end{align*}
	where (a), (b) follows from $t_0(q) < t_1(q)$. As $z_2 < 1$, $z_2^x$ is a decreasing function in $x$. Again, using $t_0 < t_1$, we know that $z_1 z_2^{t_0(q_1)} < 1$ implies $z_1 z_2^{t_1(q_1)} < 1$, which concludes that the condition of \textsc{case} $4$ is infeasible.	
\end{IEEEproof}

\section{Proof of Prop.~\ref{thm:risk_exponent}} \label{app:pf_of_risk_exp}
Let us consider an upper bound on $R_\cent^{(N)}$. Relying on the concentration property of i.i.d.~random variables, we will prove that $R_\cent^{(N)} - R_\cent^{(\infty)} \le C\cdot \exp(-cN)$ with some $C, c>0$, i.e., the Bayes risk converges to its limit value at least exponentially fast.

Consider \textsc{case} $1$ and fix $(q_\cent, q_1)$ so that the condition of \textsc{case} $1$ is satisfied. Recall the likelihood ratio test in Sec.~III-A that the fusion agent performs: After updating $q_\cent$ to $q_\cent'$ by
\begin{align*}
	\frac{q_\cent' }{1-q_\cent'} = \frac{q_\cent }{1-q_\cent} (z_1 z_2^{r_1})^N,
\end{align*}
it performs the likelihood ratio test as if $q_\cent'$ is the true prior,
\begin{align*}
	\frac{f_{Y|H}(y_\cent|1)}{f_{Y|H}(y_\cent|0)} \underset{\wh{H}_\cent = 0}{\overset{\wh{H}_\cent = 1}{\gtrless}} \frac{c_{\FA} q_\cent'}{c_{\MD}(1-q_\cent')}.
\end{align*}
Then, taking logarithm, we equivalently have
\begin{align*}
	\log \frac{f_{Y|H}(y_\cent|1)}{f_{Y|H}(y_\cent|0)} \underset{\wh{H}_\cent = 0}{\overset{\wh{H}_\cent = 1}{\gtrless}} \log \frac{c_{\FA} q_\cent}{c_{\MD}(1-q_\cent)} + N \log (z_1 z_2^{r_1}).
\end{align*}

Also, since the local decisions are i.i.d.~made as if $q_1$ is the true prior, $\wh{H}_i$ are i.i.d.~random variables from $\textsf{Bern}(t_0(q_1))$ when $H=0$, whereas $\wh{H}_i$ are from $\textsf{Bern}(t_1(q_1))$ when $H=1$. Let us take $\delta > 0$ and define two ``typical'' events $\mathcal{T}_{\delta}^0$, $\mathcal{T}_{\delta}^1$, e.g., strongly typical sets \cite{Yeung2008}:
\begin{align*}
	\mathcal{T}_{\delta}^0 &= \left\{ \wh{h}^N: \left| \frac{1}{N} \sum_{i=1}^N \wh{h}_i - t_0(q_1) \right| < \delta \right\}, \\
	\mathcal{T}_{\delta}^1 &= \left\{ \wh{h}^N: \left| \frac{1}{N} \sum_{i=1}^N \wh{h}_i - t_1(q_1) \right| < \delta \right\}.
\end{align*}
Then, the risk expression \eqref{eq:risk_expression} can be rewritten as
\begin{align*}
	R_\cent^{(N)} &= c_{\FA}\pi_0 \sum_{\wh{h}^N \in \mathcal{T}_{\delta}^0} \left( \prod_{i=1}^N p_{\wh{H}_i|H}(\wh{h}_i|0) \right) p_{\wh{H}_\cent|H, \wh{H}^N}(1|0, \wh{h}^N) \\
	&+ c_{\FA}\pi_0 \sum_{\wh{h}^N \not\in \mathcal{T}_{\delta}^0} \left( \prod_{i=1}^N p_{\wh{H}_i|H}(\wh{h}_i|0) \right) p_{\wh{H}_\cent|H, \wh{H}^N}(1|0, \wh{h}^N) \\
	&+ c_{\MD}\bar{\pi}_0 \sum_{\wh{h}^N \in \mathcal{T}_{\delta}^1} \left( \prod_{i=1}^N p_{\wh{H}_i|H}(\wh{h}_i|1) \right) p_{\wh{H}_\cent|H, \wh{H}^N}(0|1, \wh{h}^N) \\
	&+ c_{\MD}\bar{\pi}_0 \sum_{\wh{h}^N \not \in \mathcal{T}_{\delta}^1} \left( \prod_{i=1}^N p_{\wh{H}_i|H}(\wh{h}_i|1) \right) p_{\wh{H}_\cent|H, \wh{H}^N}(0|1, \wh{h}^N).
\end{align*}

Because we are considering \textsc{case} $1$ such that $z_1 z_2^{t_0(q_1)} >1$, assuming $\delta$ is small enough, $z_1 z_2^{r_1} > 1$ if $\wh{h}^N \in \mathcal{T}_\delta^0$. This implies that the decision threshold of the fusion agent after observing $\wh{h}^N$ increases linearly in $N$. From the fact that the tail probability of sub-Gaussian is upper bounded by $\exp(-t^2/2\sigma^2)$ \cite{Vershynin2018},
\begin{align}
	p_{\wh{H}_\cent|H, \wh{H}^N}(1|0, \wh{h}^N) \le \exp\left( -\frac{N^2 \Delta_0^2}{2 \sigma^2} \right), \label{eq:subGaussian_tail}
\end{align}
where $\Delta_0 \triangleq \log\left(z_1 z_2^{t_0(q_1)-\delta} \right) > 0$. Therefore, the first term is bounded by
\begin{align*}
	&c_{\FA}\pi_0 \sum_{\wh{h}^N \in \mathcal{T}_{\delta}^0} \left( \prod_{i=1}^N p_{\wh{H}_i|H}(\wh{h}_i|0) \right) p_{\wh{H}_\cent|H, \wh{H}^N}(1|0, \wh{h}^N) \\
	&\le c_{\FA}\pi_0 \sum_{\wh{h}^N \in \mathcal{T}_{\delta}^0} \left( \prod_{i=1}^N p_{\wh{H}_i|H}(\wh{h}_i|0) \right) \exp\left( -\frac{N^2 \Delta_0^2}{2 \sigma^2} \right) \\
	&\le c_{\FA}\pi_0 \exp\left( -\frac{N^2 \Delta_0^2}{2 \sigma^2} \right).
\end{align*}

To bound the second term, note that the local decisions are i.i.d.~and bounded (either $0$ and $1$). Hence, Hoeffding's bound \cite{Durrett2019} tells that
\begin{align*}
	\mathbb{P}[\wh{H}^N &\not \in \mathcal{T}_\delta^h | H=0] \le 2 \exp(-2N \delta) ~~ \textrm{for } h \in \{0,1\}.
\end{align*}
Hence the second term is bounded by
\begin{align*}
	& c_{\FA}\pi_0 \sum_{\wh{h}^N \not\in \mathcal{T}_{\delta}^0} \left( \prod_{i=1}^N p_{\wh{H}_i|H}(\wh{h}_i|0) \right) p_{\wh{H}_\cent|H, \wh{H}^N}(1|0, \wh{h}^N) \\
	&\le c_{\FA}\pi_0 \sum_{\wh{h}^N \not\in \mathcal{T}_{\delta}^0} \left( \prod_{i=1}^N p_{\wh{H}_i|H}(\wh{h}_i|0) \right) \\
	&\le c_{\FA}\pi_0 \mathbb{P} \left[\wh{H}^N \not \in \mathcal{T}_\delta^0 | H=0 \right] \le 2 c_{\FA}\pi_0 \exp(-2N \delta).
\end{align*}

We have similar bounds for the other terms and finally
\begin{align*}
	R_\cent^{(N)} &\le c_{\FA}\pi_0 \exp(-N^2 \Delta_0^2 /2) + 2 c_{\FA}\pi_0 \exp(-2N\delta) \\
	&\quad + c_{\MD}\bar{\pi}_0 \exp(-N^2 \Delta_1^2 /2) + 2 c_{\MD}\bar{\pi}_0 \exp(-2N\delta) \\
	&= O(\exp(-N \delta)),
\end{align*}
where $\Delta_1 \triangleq - \log \left( z_1 z_2^{t_1(q_1) + \delta}\right)$. Therefore, it converges to zero at least exponentially fast. This leads us to the positive constant lower bound on $\beta^*$. Other cases can be proven similarly.

\section{Proof of Thm.~\ref{thm:asymp_optimal_belief}} \label{app:pf_by_tsitsiklis}
This section shows that our model essentially makes no difference from the model in \cite{Tsitsiklis1988}, thus the result of \cite{Tsitsiklis1988} that having identical belief (that is, having identical decision threshold in LRT) is also optimal still holds for our model.

To be specific, consider a star network model given in \cite{Tsitsiklis1988} where the fusion agent knows decision rules that the local agents used. Hence, unlike \eqref{eq:belief_update}, the fusion agent can update its decision rule (or belief) based on true quantities.  It is also different from our model in that the fusion agent aggregates local decisions without a private signal, i.e., $Y_\cent = \emptyset$, and wishes to minimize probability of error, i.e., $J^{(N)} \triangleq \pi_0 p_{\wh{H}_\cent|H}(1|0) + \bar{\pi}_0 p_{\wh{H}_\cent|H}(0|1)$. Therefore, $J^{(N)}$ is a function of decision rules of the local and fusion agents. Define the optimal error exponent
\begin{align*}
	\gamma^* \triangleq \sup \left( - \lim_{N \to \infty} \frac{1}{N} \log J^{(N)} \right),
\end{align*}
where the supremum is over all possible decision rules, not necessarily identical nor LRT. Then, Tsitsiklis \cite{Tsitsiklis1988} shows that the local agents having identical decision rules indeed achieves the optimal $\gamma^*$. The statement below is adapted for binary hypothesis and decision spaces.
\begin{thm}[Thm.~1 and Sec.~II in \cite{Tsitsiklis1988}]
	The optimal error exponent $\gamma^*$ can be obtained by LRT with identical decision thresholds at the local agents, followed by LRT at the fusion agent. Specifically, the optimal local decision threshold is determined by the minimizer $\lambda^*$ of the following optimization problem\footnote{The objective function \eqref{eq:optimal_local_lam} is the (negative of) Chernoff information and also appears in the context of error exponent of discrete channels \cite{ShannonGB1967}.}: $\min_{\lambda \in \mathbb{R}} \min_{s \in [0,1]} \Lambda(\lambda, s)$ where
	\begin{align}
		&\Lambda = \log \left( p_{\wh{H}|H}^{1-s}(0|0) p_{\wh{H}|H}^s(0|1) + p_{\wh{H}|H}^{1-s}(1|0) p_{\wh{H}|H}^s(1|1) \right), \label{eq:optimal_local_lam}
	\end{align}
	and the optimal fusion agent performs the following LRT.
	\begin{align}
		\frac{p_{\wh{H}^N|H}(\wh{h}^N|1)}{p_{\wh{H}^N |H}(\wh{h}^N|0)} \underset{\wh{H}_\cent = 0}{\overset{\wh{H}_\cent = 1}{\gtrless}} \frac{\pi_0}{1-\pi_0}. \label{eq:Tsitsiklis_fusion_LRT}
	\end{align}
Finally, $\gamma^* = -\Lambda^* \triangleq -\min_{\lambda \in \mathbb{R}} \min_{s \in [0,1]} \Lambda(\lambda, s)$.
\end{thm}

First, look into \cite{Tsitsiklis1988} in detail. Since the local decision rules are all identical, $p_{\wh{H}_1|H}(\wh{h}|h) = \cdots = p_{\wh{H}_N|H}(\wh{h}|h)$. Letting $L_0 \triangleq \log \frac{p_{\wh{H}_1|H}(0|1)}{p_{\wh{H}_1|H}(0|0)}$ and $L_1 \triangleq \log \frac{p_{\wh{H}_1|H}(1|1)}{p_{\wh{H}_1|H}(1|0)}$ for brevity, \eqref{eq:Tsitsiklis_fusion_LRT} can be written equivalently as
\begin{align}
	&(\textrm{\# of $0$s in } \wh{h}^N) L_0 + (\textrm{\# of $1$s in } \wh{h}^N) L_1 \underset{\wh{H}_\cent = 0}{\overset{\wh{H}_\cent = 1}{\gtrless}} \log \frac{\pi_0}{1-\pi_0} \nonumber \\
	&\iff N(1-r_1) L_0 + N r_1 L_1 \underset{\wh{H}_\cent = 0}{\overset{\wh{H}_\cent = 1}{\gtrless}} \log \frac{\pi_0}{1-\pi_0}, \label{eq:asymp_decision1}
\end{align}
where $r_1 \triangleq \frac{\# \textrm{ of ones in } \wh{h}^N}{N}$. Although $r_1$ is a random variable, it is highly concentrated around a constant \eqref{eq:r0_asymp} so can be thought of as a constant with high probability. Then, the left side is linear in $N$ and only takes either $-\infty$ or $\infty$ as $N \to \infty$, depending on whether $H=0$ or $H=1$. However, the right side, which is the maximum a posteriori (MAP) decision threshold, is finite provided that $\pi_0$ is given. This implies that $\wh{H}_0$ is determined independently of $\pi_0$ as $N \to \infty$. Also, the optimal local decision threshold $\lambda^*$ is independent of $\pi_0$, and $\wh{H}_0$ is asymptotically independent of $\pi_0$ as well. It means that the fusion agent's decision threshold does not change the risk exponent.

Returning to our setting with the fusion agent's private observation and belief, consider the binary hypothesis testing that the fusion agent performs. Let $L_{0,[0]} \triangleq \log \frac{p_{\wh{H}_1|H}(0|1)_{[0]}}{p_{\wh{H}_1|H}(0|0)_{[0]}}$ and $L_{1,[0]} \triangleq \log \frac{p_{\wh{H}_1|H}(1|1)_{[0]}}{p_{\wh{H}_1|H}(1|0)_{[0]}}$ to emphasize the fusion agent's perception. Then,
\begin{align*}
	& \frac{f_{Y_\cent, \wh{H}^N|H}(y_\cent, \wh{h}^N|1)}{f_{Y_\cent, \wh{H}^N |H}(y_\cent, \wh{h}^N|0)} \underset{\wh{H}_\cent = 0}{\overset{\wh{H}_\cent = 1}{\gtrless}} \frac{c_\FA q_0}{c_\MD (1-q_0)} \iff \nonumber \\
	&N(1-r_1)L_{0,[0]} + N r_1 L_{1,[0]} \underset{\wh{H}_\cent = 0}{\overset{\wh{H}_\cent = 1}{\gtrless}} \\
	&\quad \quad \quad \quad \quad \quad \quad \log \frac{c_\FA q_0}{c_\MD (1-q_0)} - \log \frac{f_{Y_\cent|H}(y_\cent|1)}{f_{Y_\cent|H}(y_\cent|0)}.
\end{align*}

First, the fusion agent can exactly compute the true quantities $L_0, L_1$ if $q_0 = q_1$, which means that the probabilities in the left side agree with the quantities in \eqref{eq:asymp_decision1}. Also, the left side grows linearly in $N$. The first term in the right side is finite, so negligible in asymptotics. The second term can take a larger value than the left side which is linear in $N$, but the sub-Gaussian property guarantees that such event is with probability at most $\exp(-(cN)^2/2\sigma^2)$ \cite{Vershynin2018}, negligible in terms of the exponent. Hence, we can conclude that the right side is bounded as in the setting of \cite{Tsitsiklis1988}.

Therefore, having identical belief (thus, identical initial decision threshold as well) asymptotically minimizes the exponent of equal-cost Bayes risk $\pi_0' p_{\wh{H}_\cent|H}(1|0) + \bar{\pi}_0' p_{\wh{H}_\cent|H}(0|1)$. The only difference now is $c_\FA, c_\MD$ in $R_\cent^{(N)}$. However, $R_\cent^{(N)}$ can be thought of as a scaled version of the equal-cost risk,
\begin{align*}
	R_\cent^{(N)} &= c_\FA \pi_0 p_{\wh{H}_\cent|H}(1|0) + c_\MD \bar{\pi}_0 p_{\wh{H}_\cent|H}(0|1) \\
	&= \frac{1}{C} \left( C c_\FA \pi_0 p_{\wh{H}_\cent|H}(1|0) + C c_\MD \bar{\pi}_0 p_{\wh{H}_\cent|H}(0|1) \right) \\
	&\triangleq \frac{1}{C} \left( \pi_0' p_{\wh{H}_\cent|H}(1|0) + \bar{\pi}_0' p_{\wh{H}_\cent|H}(0|1) \right),
\end{align*}
where $C \triangleq (c_\FA \pi_0 + c_\MD \bar{\pi}_0)^{-1}$ is taken so that $C c_\FA \pi_0 + C c_\MD \bar{\pi}_0 = 1$. Since the risk exponent remains unchanged after constant scaling of risk, this is also asymptotically optimal in the risk exponent sense as $N \to \infty$. The fusion agent's decision rule has been changed from that based on $\pi_0$ to that based on $q_0 = q_1$, but this difference does not change the decision outcome. This completes the proof.

\section{Proof of Cor.~\ref{cor:Gaussian_risk_exponent}} \label{app:pf_gaussian}
\begin{figure}[t]
	\centering
	\includegraphics[width=2.4in]{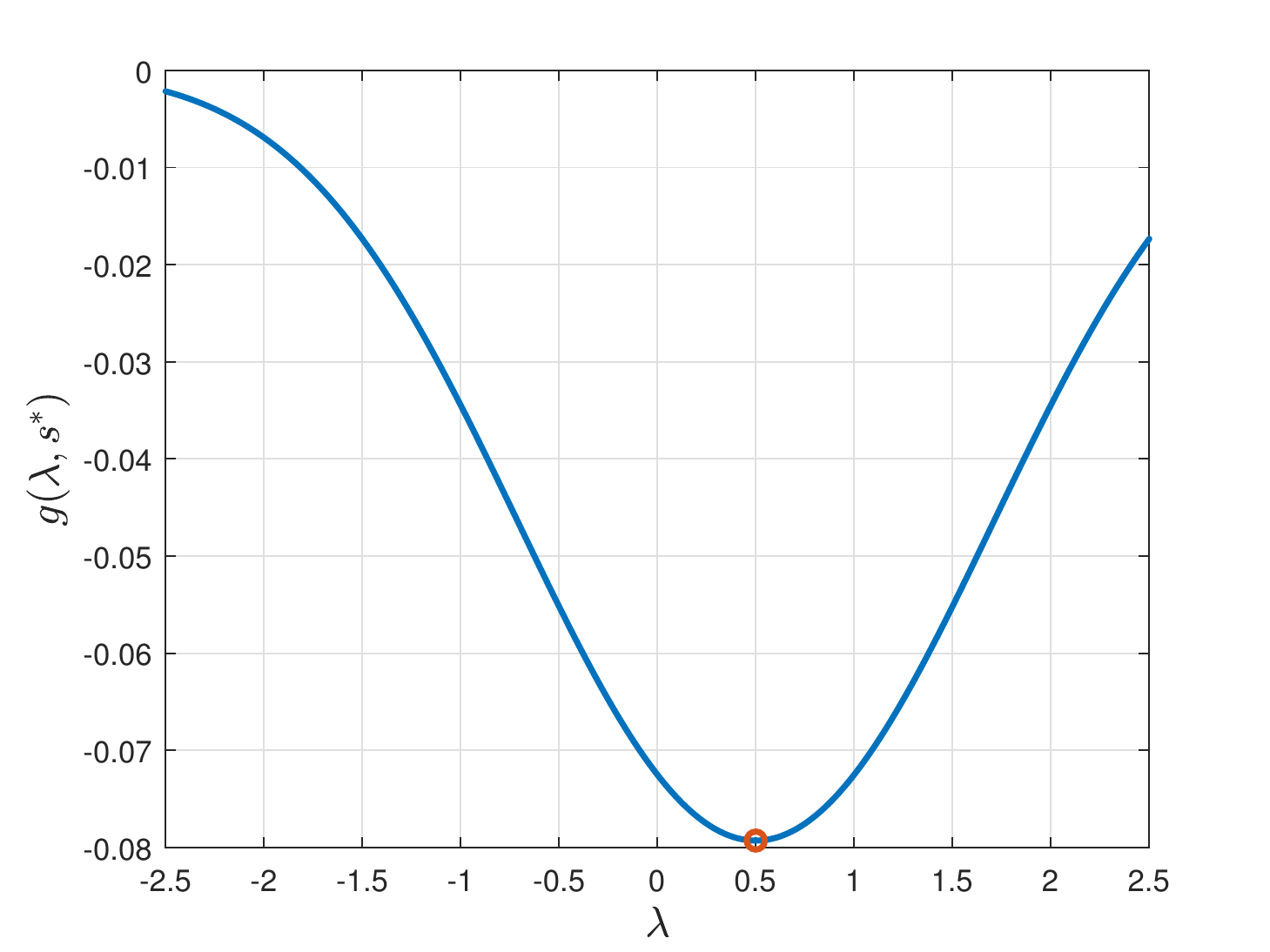}
	\caption{Plot of $g(\lambda, s^*(\lambda))$ with the minimum in red circle.}
	\label{fig:nume_sol}
\end{figure}

When a specific noise distribution is given, we can (at least numerically) compute the optimal decision threshold, corresponding to initial belief and exponent. Let us consider additive Gaussian noise. To find the optimal local threshold $\lambda^*$, recall that the objective function is continuous and differentiable in $\lambda$ and $s$, and convex in $s$, e.g., \cite{ShannonGB1967}. Hence, we first fix an arbitrary $\lambda$ and find the minimizer $s^* = s^*(\lambda)$. For notational simplicity, define $a \triangleq p_{\wh{H}|H}(0|0)$, $b \triangleq p_{\wh{H}|H}(1|1)$, and $g(\lambda, s) \triangleq \log(a^{1-s} (1-b)^s + (1-a)^{1-s} b^s)$. Find $s^*(\lambda)$ with $\lambda$ being fixed,
\begin{align*}
	&\frac{\partial}{\partial s} g(\lambda, s) = 0 \iff \\
	&-a^{1-s}(1-b)^s \log a + a^{1-s} (1-b)^s \log(1-b) \\
	&- (1-a)^{1-s} b^s \log(1-a) + (1-a)^{1-s} b^s \log b = 0.
\end{align*}
Letting $A \triangleq -\log(1-a) + \log b$, $B \triangleq \log a - \log(1-b)$, and rearranging terms with algebra, we obtain
\begin{align*}
	s^* = \log \frac{\frac{a}{1-a} + \log \frac{B}{A}}{A+B}.
\end{align*}
Now the objective is a function of only $\lambda$,
\begin{align*}
	g(\lambda, s^*(\lambda)) = \log(a^{1-s^*} (1-b)^{s^*} + (1-a)^{1-s^*} b^{s^*}).
\end{align*}
Note that $a, b, s^*$ are all dependent on $\lambda$. Due to its complexity, we are unable to analytically solve it, but the numerical solution for Gaussian noise turns out to be $\lambda^* = 0.5$ as shown in Fig.~\ref{fig:nume_sol}, which implies $q_i^* = \frac{c_\MD}{c_\FA + c_\MD}$ for all $i \in \{1,\ldots, N\}$. Hence, it is optimal for the fusion agent exactly recovers the information in the local decisions, so that $q_0^* = q_1$ as well.

The second claim is immediate from the above argument. As local decisions follow $\textsf{Bern}(Q(0.5/\sigma))$ if $H=0$ and $\textsf{Bern}(Q(-0.5/\sigma))$ if $H=1$, this can be cast into Bayesian binary hypothesis testing with i.i.d.~observations. The fusion agent's private observation and the true prior are negligible as discussed. Therefore, the Chernoff information is the optimal risk exponent, which is the (negative of) objective function \eqref{eq:optimal_local_lam}. As in Fig.~\ref{fig:nume_sol}, the Chernoff information is
\begin{align*}
	\beta^* &= C(\textsf{Bern}(Q(0.5/\sigma)), \textsf{Bern}(Q(-0.5/\sigma))) \\
	&\approx 0.0793 ~~~ \textrm{if $\sigma = 1$.}
\end{align*}

\bibliographystyle{IEEEtran}
\bibliography{abrv,conf_abrv,DW_lib,lrv_lib}

\end{document}